\documentclass[
journal
]{IEEEtran}

\usepackage{amsmath}
\usepackage{amssymb}
\usepackage{amsfonts}
\usepackage{amsthm}
\usepackage{array}
\usepackage{cite}
\usepackage{float}

\newcommand{\bcomment}[1]{}
\newcommand{\TODO}[1]{}

\newcommand{\R}{\mathbb{R}}
\newcommand{\N}{\mathbb{N}}
\newcommand{\Z}{\mathbb{Z}}

\newcommand{\E}{\mathbb{E}}

\newcommand{\twobytwo}[4]{\begin{pmatrix} #1 & #2 \\ #3 & #4 \end{pmatrix}}

\newcommand{\bfp}{\mathbf{p}}

\newcommand{\bfx}{\mathbf{x}}

\newcommand{\mcE}{\mathcal{E}}

\newcommand{\eq}{\mathrel{\phantom{=}}}

\newcommand{\one}{\mathbf{1}}

\newcommand{\multisetds}[2]{\bigg(\kern-.4em\binom{#1}{#2}\kern-.4em\bigg)}
\newcommand{\multisetin}[2]{\big(\kern-.3em\binom{#1}{#2}\kern-.3em\big)}
\newcommand{\multisetix}[2]{\left(\kern-.2em\binom{#1}{#2}\kern-.2em\right)}

\newcommand{\deq}{\triangleq}

\newcommand{\eql}[1]{\overset{(#1)}{=}}
\newcommand{\leql}[1]{\overset{(#1)}{\leq}}
\newcommand{\geql}[1]{\overset{(#1)}{\geq}}
\newcommand{\propl}[1]{\overset{(#1)}{\propto}}

\DeclareMathOperator*{\argmin}{argmin}
\DeclareMathOperator*{\argmax}{argmax}

\usepackage{tikz}

\newtheorem{theorem}{Theorem}

\newtheorem{lemma}{Lemma}[section]
\newtheorem{corollary}{Corollary}
\newtheorem{definition}{Definition}
\newtheorem{example}{Example}

\newcommand{\ttl}{\tilde{t}}
\newcommand{\ntl}{\tilde{n}}

\newcommand{\mtl}{\widetilde{m}}
\newcommand{\Mtl}{\widetilde{M}}
\newcommand{\Atl}{\widetilde{A}}
\newcommand{\Stl}{\widetilde{\mathcal{S}}}
\newcommand{\ER}{Erd\H{o}s-R\'{e}nyi }
\title{Exact alignment recovery for correlated \ER graphs}
\author{Daniel~Cullina,~\IEEEmembership{Member,~IEEE}
        and Negar~Kiyavash,~\IEEEmembership{Senior Member,~IEEE}%
  \thanks{The material in this paper was presented in part at the ACM SIGMETRICS conference, Antibes Juan-les-Pins, France, June 2016 \cite{cullina_improved_2016}.}%
\thanks{Daniel Cullina is with the Department of Electrical, Princeton University, Princeton, New Jersey 08540 (email: dcullina@princeton.edu). }%
\thanks{Negar Kiyavash is with the Department of Electrical and Computer Engineering, the Department of Industrial and Enterprise Systems Engineering, and the Coordinated Science Laboratory, University of Illinois at Champaign-Urbana, Urbana, Illinois 61801 (email: kiyavash@illinois.edu). }%
}
\begin{document}

\maketitle

\begin{abstract}
  We consider the problem of perfectly recovering the vertex correspondence between two correlated \ER (ER) graphs on the same vertex set.
  The correspondence between the vertices can be obscured by randomly permuting the vertex labels of one of the graphs.
  We determine the information-theoretic threshold for exact recovery, i.e. the conditions under which the entire vertex correspondence can be correctly recovered given unbounded computational resources.
\end{abstract}

Graph alignment is the problem finding a matching between the vertices of the two graphs that matches, or aligns, many edges of the first graph with edges of the second graph.
Alignment is a generalization of graph isomorphism recovery to non-isomorphic graphs.
Graph alignment can be applied in the deanonymization of social networks, the analysis of protein interaction networks, and computer vision.
Narayanan and Shmatikov successfully deanonymized an anonymized social network dataset by graph alignment with a publicly available network \cite{narayanan_de-anonymizing_2009}.
In order to make privacy guarantees in this setting, it is necessary to understand the conditions under which graph alignment recovery is possible.

We consider graph alignment for a randomized graph-pair model.
This generation procedure creates a ``planted'' alignment: there is a ground-truth relationship between the vertices of the two graphs.
Pedarsani and Grossglauser~\cite{pedarsani_privacy_2011} were the first to approach the problem of finding information-theoretic conditions for alignment recovery.
They established conditions under which exact recovery of the planted alignment is possible,
The authors improved on these conditions and also established conditions under which exact recover is impossible~\cite{cullina_improved_2016}.
In this paper, we close the gap between these results and establish the precise threshold for exact recovery in sparse graphs.
As a special case, we recover a result of Wright~\cite{wright_graphs_1971} about the conditions under which an \ER graph has a trivial automorphism group.

\section{Model}
\label{section:model}
\subsection{The alignment recovery problem}
\label{section:problem}
We consider the following problem.
There are two correlated graphs $G_a$ and $G_b$, both on the vertex set $[n] = \{0,1,\ldots,n-1\}$.
By correlation we mean that for each vertex pair $e$, presence or absence of $e \in E(G_a)$, or equivalently the indicator variable $G_a(e)$, provides some information about $G_b(e)$.
The true vertex labels of $G_a$ are removed and replaced with meaningless labels.
We model this by applying a uniformly random permutation $\Pi$ to map the vertices of $G_a$ to the vertices of its anonymized version.
The anonymized graph is $G_c$, where $G_c(\{\Pi(i),\Pi(j)\}) = G_a(\{i,j\})$ for all $i,j \in [n]$, $i \neq j$.
The original vertex labels of $G_b$ are preserved and $G_c$ and $G_b$ are revealed.
We would like to know under what conditions it is possible to discover the true correspondence between the vertices of $G_a$ and the vertices of $G_b$.
In other words, when can the random permutation $\Pi$ be exactly recovered with high probability?

In this context, an achievability result demonstrates the existence of an algorithm or estimator that exactly recovers $\Pi$ with high probability.
A converse result is an upper bound on the probability of exact recovery that applies to any estimator.

\subsection{Correlated \ER graphs}
To fully specify this problem, we need to define a joint distribution over $G_a$ and $G_b$.
In this paper, we will focus on \ER (ER) graphs.
We discuss some of the advantages and drawbacks of this model in Section~\ref{section:related-work}.

We will generate correlated \ER graphs as follows.
Let $G_a$ and $G_b$ be graphs on the vertex set $[n]$.
We will think of $(G_a, G_b)$ as a single function with codomain $\{0,1\}^2$: $(G_a,G_b)(e) = (G_a(e), G_b(e))$.
The random variables $(G_a,G_b)(e)$, $e \in \binom{[n]}{2}$, are i.i.d. and
\begin{equation*}
  (G_a,G_b)(e) = 
  \begin{cases}
    (1,1) & \text{w.p. } p_{11} \\
    (1,0) & \text{w.p. } p_{10} \\
    (0,1) & \text{w.p. } p_{01} \\
    (0,0) & \text{w.p. } p_{00}.
  \end{cases}
\end{equation*}
Call this distribution $ER(n,\bfp)$, where $\bfp = (p_{11},p_{10},p_{01},p_{00})$.
Note that the marginal distributions of $G_a$ and $G_b$ are \ER and so is the distribution of the intersection graph $G_a \wedge G_b$: $G_a \sim ER(n,p_{10}+p_{11})$, $G_b \sim ER(n,p_{01}+ p_{11})$, and $G_a \wedge G_b \sim ER(n,p_{11})$.

When $p_{11} > (p_{10}+p_{11})(p_{01}+p_{11})$, we say that the graphs $G_a$ and $G_b$ have positive correlation.
Observe that
\bcomment{
\begin{align}
  &\eq 1 - \frac{(p_{10}+p_{11})(p_{01}+p_{11})}{p_{11}}\nonumber\\ 
  &= 1 - p_{11} - p_{10} - p_{01} - \frac{p_{10}p_{01}}{p_{11}}  \nonumber\\
  &= p_{00}\left(1 - \frac{p_{10}p_{01}}{p_{11}p_{00}}\right),
\end{align}}
\begin{equation*}
p_{11} - (p_{10}+p_{11})(p_{01}+p_{11}) = p_{11}p_{00} - p_{01}p_{10}
\end{equation*}
so $p_{11}p_{00} > p_{10}p_{01}$ is an equivalent, more symmetric condition for positive correlation.
In this paper, we only address the case of positive correlation.

\subsection{Results}
All of the results concern the following setting.
We have $(G_a,G_b) \sim ER(n,\bfp)$, $\Pi$ is a uniformly random permutation of $[n]$ independent of $(G_a,G_b)$, and $G_c$ is the anonymization of $G_a$ by $\Pi$ as described in Section~\ref{section:problem}.
Our main result is the following.
\begin{theorem}
  \label{thm:main}
  Let $\bfp$ satisfy the conditions
\begin{align}
  p_{11} &\geq \frac{\log n + \omega(1)}{n}. \label{p11-main}\\
  p_{11} &\leq \mathcal{O}{\left(\frac{1}{\log n}\right)} \label{p11-sparse}\\
  p_{01} +p_{10} &\leq \mathcal{O}{\left(\frac{1}{\log n}\right)} \label{p-sparse}\\
  \frac{p_{01}p_{10}}{p_{11}p_{00}} &\leq \mathcal{O}{\left(\frac{1}{(\log n)^3}\right)} \label{p-corr}
\end{align}
Then there is an estimator for $\Pi$ given $(G_c,G_b)$ that is correct with probability $1 - o(1)$.
\end{theorem}

Together, conditions \eqref{p11-sparse} and \eqref{p-sparse} force $G_a$ and $G_b$ to be mildly sparse.
Condition \eqref{p-corr} requires $G_a$ and $G_b$ to have nonnegligible positive correlation.

We also have a converse bound.
\newcommand{\thmconverse}{%
If $\bfp$ satisfies
  \begin{align*}
    p_{11} &\leq \frac{\log n - \omega(1)}{n}\\
    \frac{p_{01}p_{10}}{p_{11}p_{00}} &< 1,
  \end{align*}
  then any estimator for $\Pi$ given $(G_c,G_b)$ is correct with probability $o(1)$.
}
\begin{theorem}[Converse bound]
  \label{thm:converse}
  \thmconverse
\end{theorem}
Theorem~\ref{thm:converse} does not require conditions \eqref{p11-sparse} or \eqref{p-sparse} and requires only a weaker version of \eqref{p-corr}.
In the regime where Theorem~\ref{thm:main} applies, Theorem~\ref{thm:converse} matches it.
Theorem~\ref{thm:converse} was originally proved by the authors in \cite{cullina_improved_2016}.
The proof is short compared to Theorem~\ref{thm:main} and it is included in Section~\ref{section:converse}.

\newcommand{\thmachone}{%
  If $\bfp$ satisfies  
  \begin{align*}
    (\sqrt{p_{11}p_{00}} - \sqrt{p_{01}p_{10}})^2 &\geq \frac{2 \log n + \omega(1)}{n}\\
    \frac{p_{01}p_{10}}{p_{11}p_{00}} &< 1.
  \end{align*}
  then there is an estimator for $\Pi$ given $(G_c,G_b)$ that is correct with probability $1 - o(1)$.
}

A second achievability theorem applies without conditions \eqref{p11-sparse}, \eqref{p-sparse}, and \eqref{p-corr}.
This requires condition \eqref{p11-main} to be strengthened.
\begin{theorem}
  \label{thm:ach-one}
  \thmachone
\end{theorem}
Theorem~\ref{thm:ach-one} was also originally proved in \cite{cullina_improved_2016}.
In this paper, it appears as an intermediate step in the proof of Theorem~\ref{thm:main}.

\section{Preliminaries}
\subsection{Notation}
Throughout, we use capital letters for random objects and lower case letters for fixed objects.

For a graph $g$, let $V(g)$ and $E(g)$ be the node and edge sets respectively. 
Let $[n]$ denote the set $\{0,\cdots,n-1\}$.
All of the $n$-vertex graphs that we consider will have vertex set $[n]$.
We will always think of a permutation as a bijective function $[n] \to [n]$.
The set of permutations of $[n]$ under the binary operation of function composition forms the group $S_{n}$.

We denote the collection of all two element subsets of $[n]$ by $\binom{[n]}{2}$.
The edge set of a graph $g$ is $E(g) \subseteq \binom{[n]}{2}$.

Represent a labeled graph on the vertex set $[n]$ by its edge indicator function $g : \binom{[n]}{2} \to [2]$.
The group $S_n$ has an action on $\binom{[n]}{2}$.
We can write the action of the permutation $\pi$ on the graph $g$ as the composition of functions $g \circ l(\pi)$, where $l(\pi)$ is the lifted version of $\pi$:
\begin{IEEEeqnarray*}{rCl}
  l(\pi) &:& \binom{[n]}{2} \to \binom{[n]}{2}\\
  && \{i,j\} \mapsto \{\pi(i), \pi(j)\}.
\end{IEEEeqnarray*}
Thus $G_c = G_a \circ l(\Pi^{-1})$.
Whenever there is only a single permutation under consideration, we will follow the convention $\tau = l(\pi)$.

For a generating function in the formal variable $z$, $[z^j]$ is the coefficient extraction operator:
\[
  [z^j] a(z) = [z^j] \sum_i a_i z^i = a_j .
\]
When $z$ is a matrix of numbers or formal variables and $k$ is a matrix of numbers, both indexed by $\mathcal{S} \times \mathcal{T}$, we use the notation
\[
  z^k = \prod_{i \in \mathcal{S}} \prod_{j \in \mathcal{T}}  z_{i,j}^{k_{i,j}}
\]
for compactness.

\subsection{Graph statistics}
Recall that we consider a graph on $[n]$ to be a $[2]$-labeling of the set of vertex pairs $\binom{[n]}{2}$.
The following quantities have clear interpretations for graphs, but we define them more generally for reasons that will become apparent shortly.
\begin{definition}
  For a set $\mathcal{S}$ and a pair of binary labelings $f_a,f_b : \mathcal{S} \to [2]$, define the \emph{type} 
  \begin{align*}
    \mu(f_a,f_b) &\in \N^{[2] \times [2]}\\ 
    \mu(f_a,f_b)_{ij} &= \sum_{e \in \mathcal{S}} \one\{(f_a, f_b)(e)=(i,j)\} .
  \end{align*}

  The \emph{Hamming distance} between $f_a$ and $f_b$ is
  \begin{align*}
    \Delta(f_a,f_b)
    &= \sum_{e \in \mathcal{S}} \one\{ f_a(e) \neq f_b(e)\}\\
    &= \mu(f_a,f_b)_{01} + \mu(f_a,f_b)_{10} ,
  \end{align*}
  For a permutation $\tau : \mathcal{S} \to \mathcal{S}$, define 
  \[
    \delta(\tau;f_a,f_b) = \frac{1}{2}\left(\Delta(f_a \circ \tau, f_b) - \Delta(f_a, f_b)\right).
  \]
\end{definition}

In the particular case of graphs (where $\mathcal{S} = \binom{[n]}{2}$ and $\tau = l(\pi)$), $\Delta(g_a,g_b)$ is the size of the symmetric difference of the edge sets, $|E(g_a \vee g_b)| - |E(g_a \wedge g_b)|$.
The quantity $\delta$ is central to both our converse and our achievability arguments (as well as the achievability proof of Pedarsani and Grossglauser \cite{pedarsani_privacy_2011}).
When $f_a$ and $f_b$ are graphs on $[n]$ and $\pi$ is a permutation of $[n]$, $\delta(l(\pi);f_a,f_b)$ is the difference in matching quality between the permutation $\pi$ and the identity permutation.

\begin{lemma}
  \label{lemma:type-diff}
  Let $f_a,f_b : \mathcal{S} \to [2]$. Then there is some $i \in \Z$ such that
  \[
    \mu(f_a \circ \tau, f_b) - \mu(f_a, f_b) = 
    \twobytwo{-i}{i}{i}{-i}
  \]
  and $i = \delta(\tau;f_a,f_b)$. 
\end{lemma}
\begin{IEEEproof}
  Let $k = \mu(f_a, f_b)$, and $k' = \mu(f_a \circ \tau, f_b)$.
  Let $\one$ be the vector of all ones.
  We have $k\one = k'\one$ because both vectors give the distribution of symbols in $f_a$.
  Similarly $\one^Tk = \one^Tk'$.
  The matrix $k'-k$ has integer entries, so it must have the claimed form for some value of $i$.
  Finally, 
  \begin{align*}
    i
    &= \frac{1}{2}((k'_{01} + k'_{10}) - (k_{01} + k_{10})) \\
    &= \frac{1}{2}(\Delta(G_a \circ \tau, G_b) - \Delta(G_a, G_b))\\
    &=\delta(\tau; f_a, f_b). \IEEEQEDhere 
  \end{align*}
\end{IEEEproof}

\subsection{MAP estimation}
The maximum a posteriori (MAP) estimator for this problem can be derived as follows.
In the following lemma we will be careful to distinguish graph-valued random variables from fixed graphs: we name the former with upper-case letters and the latter with lower-case.
\begin{lemma}
  \label{lemma:posterior}
  Let $(G_a,G_b) \sim ER(n,\bfp)$, let $\Pi$ be a uniformly random permutation of $[n]$, and let $G_c = G_a \circ l(\Pi^{-1})$.
  Then 
  \begin{equation*}
    P[\Pi = \pi|(G_c,G_b) = (g_c,g_b)] \propto \left(\frac{p_{10} p_{01}}{p_{11} p_{00}}\right)^i
  \end{equation*}
  where $i = \frac{1}{2}\Delta(g_c \circ l(\pi), g_b)$.
\end{lemma}
\begin{IEEEproof}
  \begin{align*}
    &\eq P[\Pi = \pi | (G_c, G_b) = (g_c,g_b)] \\
    &\propl{a} P[\Pi = \pi, (G_c, G_b) = (g_c,g_b)] \\
    &\eql{b} P[\Pi = \pi, (G_a, G_b) = (g_c \circ l(\pi),g_b)] \\
    &\propl{c} P[(G_a, G_b) = (g_c \circ l(\pi),g_b)] \\
    &\eql{d} p^{\mu(g_c \circ l(\pi),g_b)} \\
    &\propl{e} p^{\mu(g_c \circ l(\pi),g_b) - \mu(g_c,g_b)} \left(\frac{p_{01} p_{10}}{p_{00} p_{11}}\right)^{\frac{1}{2}\Delta(g_c,g_b)}\\    
    &\eql{f} \left(\frac{p_{01} p_{10}}{p_{00} p_{11}}\right)^{\frac{1}{2}\Delta(g_c \circ l(\pi),g_b)}
  \end{align*}
  where in $(a)$ we multiply by the constant $P[(G_c, G_b) = (g_c,g_b)]$,
    in $(b)$ we apply the relationship $G_c = G_a \circ l(\Pi^{-1})$,
    and in $(c)$ we use the independence of $(G_a,G_b)$ from $\Pi$ and the uniformity of $\Pi$.
    In $(d)$ we apply the definition of the distribution of $(G_a,G_b)$,
    in $(e)$, we divide by a constant that does not depend on $\Pi$,
    and in $(f)$ we use Lemma~\ref{lemma:type-diff}.
\end{IEEEproof}

When we have $p_{01}p_{10} < p_{11}p_{00}$, the MAP estimator is 
\begin{align*}
  \hat{\Pi}
  &= \argmax_{\hat{\pi}} P[\Pi = \hat{\pi}|(G_c,G_b)]\\
  &= \argmin_{\hat{\pi}} \Delta(G_c \circ l(\hat{\pi}),G_b).
\end{align*}

\bcomment{
Rather than tracking the actual values of $\Pi$ and $\hat{\pi}$, we focus on $\Pi^{-1} \circ \hat{\pi}$, the ``difference'' between the estimator and the correct answer.
Thus we look at
\begin{align*}
  \hat{\pi} &\mapsto \Delta(G_a \circ l(\Pi^{-1} \circ \hat{\pi}),G_b)
\end{align*}

\begin{align*}
  \pi &\mapsto \Delta(G_a \circ l(\pi),G_b)
\end{align*}}

The permutation $\hat{\pi} = \Pi$ achieves an alignment score of $\Delta(G_a,G_b)$.
Although $\Delta(G_a,G_b)$ is unknown to the estimator, we can analyze its success by considering the distribution of
\[
  \Delta(G_a \circ l(\Pi^{-1} \circ \hat{\pi}),G_b) - \Delta(G_a,G_b) = \delta(l(\Pi^{-1} \circ \hat{\pi});G_a,G_b).
\]

Let 
\begin{align*}
  \mathcal{Q}
  &= \{\pi \in S_n : \Delta(G_a \circ l(\pi), G_b) \leq \Delta(G_a,G_b)\}\\
  &= \{\pi \in S_n : \delta(l(\pi); G_a, G_b) \leq 0\},
\end{align*}
the set of permutation that give alignments of $G_a$ and $G_b$ that are at least as good as the true permutation.
The identity permutation $id$ achieves $\delta(l(id); G_a,G_b) = 0$, so it is always in $\mathcal{Q}$ by definition.

Let $\eta(G_a,G_b)$ be the probability of success of the MAP estimator conditioned on the generation of the graph pair $(G_a,G_b)$.
When $id$ is not minimizer of $\Delta(G_c \circ l(\pi), G_b)$, i.e. there is some $\pi$ such that $\delta(l(\pi);G_a,G_b) < 0$, $\eta = 0$.
When $id$ achieves the minimum, $\eta = 1/|\mathcal{Q}|$.

The converse argument uses the fact the overall probability of success is at most $\E[1/|\mathcal{Q}|]$.

The achievability arguments use the fact the overall probability of error is at most
\[
  P[|\mathcal{Q}| \geq 1] \leq \E[|\mathcal{Q}| - 1]
\]
or equivalently
\[
  P\bigg[\bigvee_{\pi \neq id} (\pi \in \mathcal{Q})\bigg] \leq \sum_{\pi \neq id} P[\pi \in \mathcal{Q}].
\]
Here we have applied linearity of expectation on the indicators for $\pi \in \mathcal{Q}$ or equivalently the union bound on these events.

\subsection{Cycle decomposition and the nontrivial region}
The cycle decompositions of the permutations $\pi$ and $\tau = \ell(\pi)$ play a crucial role in the distribution of $\delta(\tau;G_a,G_b)$.
For a vertex set $[n]$ and a fixed $\tau$, define $\Stl$, the \emph{nontrivial region} of the graph, to be the vertex pairs that are not fixed points of $\tau$, i.e. $\Stl = \left\{e \in \binom{[n]}{2} : \tau(e) \neq e \right\}$.
We will mark quantities and random variables associated with the nontrivial region with tildes.

Recall that $n$ is the number of vertices and let $\ntl$ be the number of vertices that are not fixed points of $\pi$. 
Let $t = \binom{n}{2}$, let $\ttl = |\Stl|$, and let $t_i$ be the number of vertex pairs in cycles of length $i$.
Then $\ttl = t - t_1$.

The expected value of $\delta(\tau;G_a,G_b)$ depends only on the size of the nontrivial region.
\begin{lemma}
  \label{lemma:mean}
  $\E[\delta(\tau;G_a,G_b)] = \ttl(p_{00} p_{11} - p_{01} p_{10})$.
\end{lemma}
\begin{IEEEproof}
  Let $\mathcal{S} = \binom{[n]}{2}$.
  Using the alternative expression for $\delta(\tau;G_a,G_b)$ from Lemma~\ref{lemma:type-diff}, we have
  \begin{align*}
    &\eq\E[\delta(\tau;G_a,G_b)]\\
    &= \E[\mu(G_a,G_b)_{11} - \mu(G_a \circ \tau, G_b)_{11}] \\
    &= \sum_{e \in \mathcal{S}} P(G_a(e) = G_b(e) = 1) - P(G_a(\tau(e)) = G_b(e) = 1)\\
    &= \sum_{e \in \Stl} p_{11} - (p_{10}+p_{11})(p_{01}+p_{11})\\
    &= \ttl(p_{00} p_{11} - p_{01} p_{10}). \qedhere
  \end{align*}  
\end{IEEEproof}

Let $M = \mu(G_a,G_b)_{11}$, which is the number of edges in $G_a \wedge G_b$.
Let $\Mtl$ be the number of edges in $G_a \wedge G_b$ in the nontrivial region, i.e. $|E(G_a \wedge G_b) \cap \Stl|$.
When $(G_a,G_b) \sim ER(\bfp,n)$, the events $(G_a,G_b)(e) = (1,1)$ for $e \in \binom{[n]}{2}$ are independent and occur with probability $p_{11}$, so both $M$ and $\Mtl$ have binomial distributions.
Conditioned on $M$, $\Mtl$ has a hypergeometric distribution.

We use the following notation for binomial and hypergeometric distributions.
Each of these probability distributions models drawing from a pool of $n$ items, $b$ of which are marked.
If we take $a$ samples without replacement, the number of marked items drawn has the hypergeometric distribution $\text{Hyp}(n,a,b)$.
If we take $a$ samples with replacement, the number of marked items drawn has a binomial distribution $\text{Bin}(n,a,b)$.
Thus
\begin{align}
  M &\sim \text{Bin}(t,p_{11},1)\\
  \Mtl &\sim \text{Bin}(\ttl,p_{11},1)\\
  \Mtl|M=m &\sim \text{Hyp}(\ttl,m,t).\label{mtl-gf-hyper}
\end{align}

Hypergeometric and binomial random variables have the following generating functions:
\begin{align*}
  \text{Hyp}(a,b,n;z) 
  &\deq \frac{[x^a y^b] (1 + x + y + xyz)^n}{[x^a y^b] (1 + x)^n(1 + y)^n} \\
  \text{Bin}(a,b,n;z) 
  &\deq \left(1 - \frac{b}{n} + \frac{b}{n}z\right)^a
\end{align*}

Observe that $\text{Hyp}(a,b,n;z)$ is symmetric in $a$ and $b$.
Additionally $\text{Hyp}(a,b,n;z) = z^a \text{Hyp}(a,n-b,n;z^{-1})$ because the number of marked balls that are drawn is equal to the number of draws minus the number of unmarked balls drawn.
For the same reason, $\text{Bin}(a,b,n;z) = z^a \text{Bin}(a,n-b,n;z^{-1})$.

\bcomment{
  There are two alternate expressions for the hypergeometric generating function
  \begin{align*}
    \text{Hyp}(a,b,n;z)
    &= \frac{[x^a] (1+x)^{n-b}(1+xz)^b}{[x^a](1+x)^n} \\
    &= \frac{[y^b] (1+y)^{n-a}(1+yz)^a}{[y^b](1+y)^n} 
  \end{align*}
}

\bcomment{
explain cycle decomp
explain nontrivial cycles
explain tilde notat

\begin{align*}
  [n]                &= V(G_a) = V(G_b)\\
  \ntl               &= |\{j \in [n] : \pi(j) \neq j \}|\\
  t                  &= \binom{n}{2} = \left|\binom{[n]}{2}\right|\\
  \Stl               &= \{e \in \mathcal{S} : \tau(e) \neq e \}\\
  \ttl               &= |\Stl|
                     = t - t_1
                     = \sum_{i \geq 2} t_i\\
  E(G_a \wedge G_b) &= \{e \in \mathcal{S} : (G_a,G_b)(e) = (1,1)\}\\
  M                  &= \mu(G_a,G_b)_{11}\\
                     &= |E(G_a \wedge G_b)| \\
  \Mtl               &= |E(G_a \wedge G_b) \cap \Stl| \\
\end{align*}

\TODO{
  The permutation tau acts on vertex pairs
  it partitions these into cycle
}
}

\subsection{Proof outline}
Both of our achievability proofs have the following broad structure.
\begin{itemize}
\item Use a union bound over the non-identity permutations and estimate $P[\delta(l(\pi),G_a,G_b) \leq 0]$, where $\pi$ is fixed and $(G_a,G_b)$ are random.
\item For a fixed $\pi$, examine the cycle decomposition and relate $\delta(l(\pi))$ to $\delta(l(\pi'))$, where $\pi'$ has the same number of fixed points as $\pi$ but all nontrivial cycles have length two. This is summarized in Theorem~\ref{thm:l-cycle-two-cycle}.
\item Use large deviations methods to bound the lower tail of $\delta(l(\pi'))$. This is done in Theorem~\ref{thm:opt-z}.
\end{itemize}

Our first achievability result, Theorem~\ref{thm:ach-one}, comes from applying Theorem~\ref{thm:opt-z} in a direct way.
This requires no additional assumptions on $\bfp$ but does not match the converse bound.
If $G_a \wedge G_b$ has no edges, every permutation is in $\mathcal{Q}$ and the union bound is extremely loose.
When $p_{11} = \frac{c \log n}{n}$, the probability that $G_a \wedge G_b$ has no edges is
\[
  (1-p_{11})^t \approx \exp(-tp_{11}) = \exp\left(-\frac{c}{2} (n-1) \log n\right).
\]
When $c \leq 2$, this probability is larger than $1/n!$, so the union bound on the error probability becomes larger than one.

To overcome this, in the proof of Theorem~\ref{thm:ach-two} we condition on $M = \mu(G_a,G_b)_{11}$ before applying Theorem~\ref{thm:opt-z}.
It is more difficult to apply Theorem~\ref{thm:opt-z} to $G_a,G_b|M$.
In particular, $\Mtl$, the number of edges of the intersection graph in nontrivial cycles of $\tau$, now has a hypergeometric distribution $\text{Hyp}(\ttl,m,t)$ rather than a binomial distribution $\text{Bin}(\ttl,p_{11},1)$.
One way to analyze the tail of a hypergeometric random variable is to look at the binomial random variable with the same mean and number of samples.
This idea is formalized in Lemma~\ref{lemma:hyper}.
Moving from $\text{Hyp}(\ttl,m,t)$ to $\text{Bin}(\ttl,m,t)$ would effectively undo our conditioning on $M$.
For the most important values of $\ttl$ and the typical values of $m$, we have $\ttl << m$.
Thus we exploit the symmetry of the hypergeometric distribution ($\text{Hyp}(\ttl,m,t) = \text{Hyp}(m,\ttl,t)$) and replace $\text{Hyp}(m,\ttl,t)$ with $\text{Bin}(m,\ttl,t)$, which is more concentrated than $\text{Bin}(\ttl,p_{11},1)$.

\subsection{Related Work}
\TODO{}
\label{section:related-work}
In the perfect correlation limit, i.e. $p_{01} = p_{10} = 0$, we have $G_a = G_b$.
In this case, the size of the automorphism group of $G_a$ determines whether it is possible to recover the permutation applied to $G_a$.
This is because the composition of an automorphism with the true matching gives another matching with no errors.
Whenever the automorphism group of $G_a$ is nontrivial, it is impossible to exactly recover the permutation with high probability.
We will return to this idea in Section~\ref{section:converse} in the proof of the converse part of Theorem~\ref{thm:main}.
Wright established that for $\frac{\log n + \omega(1)}{n} \leq p \leq 1 - \frac{\log n + \omega(1)}{n}$, the automorphism group of $G \sim ER(n,p)$ is trivial with probability $1 - o(1)$ and that for $p \leq \frac{\log n - \omega(1)}{n}$, it is nontrivial with probability $1-o(1)$\cite{wright_graphs_1971}.
In fact, he proved a somewhat stronger statement about the growth rate of the number of unlabeled graphs that implies this fact about automorphism groups.
Thus our Theorem~\ref{thm:main} and Theorem~\ref{thm:converse} extend Wright's results.
Bollob\'{a}s later provided a more combinatorial proof of this automorphism group threshold function \cite{bollobas_distinguishing_1982}.
The methods we use are closer to those of Bollob\'{a}s.

Onaran, Garg, and Erkip investigated the effect of community structure in a stochastic block model on the information-theoretic threshold for recovery of a graph matching \cite{onaran_optimal_2016}.
If the networks being aligned correspond to two distinct online services, it is unlikely that the user populations of the services are identical.
Kazemi, Yartseva, and Grossglausser investigate alignment recovery of correlated graphs on overlapping but not identical vertex sets \cite{kazemi_when_2015}.
They determine that the information-theoretic penalty for imperfect overlap between the vertex sets of $G_a$ and $G_b$ is relatively mild.
This regime is an important test of the robustness of alignment procedures.

Some practical recovery algorithms start by attempting to locate a few seeds.
From these seeds, the graph matching is iteratively extended.
Algorithms for the latter step can scale efficiently.
Narayanan and Shmatikov were the first to apply this method \cite{narayanan_de-anonymizing_2009}.
They evaluated their performance empirically on graphs derived from social networks.

More recently, there has been some work evaluating the performance of this type of algorithm on graph inputs from random models.
Yartseva and Grossglauser examined a simple percolation algorithm for growing a graph matching \cite{yartseva_performance_2013}.
They find a sharp threshold for the number of initial seeds required to ensure that final graph matching includes every vertex.
The intersection of the graphs $G_a$ and $G_b$ plays an important role in the analysis of this algorithm.
Kazemi, Hassani, and Grossglauser extended this work and investigated the performance of a more sophisticated percolation algorithm \cite{kazemi_growing_2015}.
Shirani, Garg, and Erkip determine the number of seeds required by a typicality matching algorithm \cite{shirani_seeded_2017}.

\subsection{Subsampling model}
\bcomment{Also define the marginal probabilities for $G_a$ and $G_b$:
\begin{IEEEeqnarray*}{rCl}
  p_{1*} &=& p_{11} + p_{10}\\ 
  p_{0*} &=& p_{01} + p_{00}\\ 
  p_{*1} &=& p_{11} + p_{01}\\ 
  p_{*0} &=& p_{10} + p_{00}.
\end{IEEEeqnarray*}}

\label{subsection:subsampling}
Pedarsani and Grossglauser \cite{pedarsani_privacy_2011} introduced the following generative model for correlated \ER (ER) graphs.
Essentially the same model was used in \cite{ji_structural_2014,ji_your_2015}.
Let $H$ be an ER graph on $[n]$ with edge probability $r$.
Let $G_a$ and $G_b$ be independent random subgraphs of $H$ such that each edge of $H$ appears in $G_a$ and in $G_b$ with probabilities $s_a$ and $s_b$ respectively.
We will refer to this as the \emph{subsampling model}.
The $s_a$ and $s_b$ parameters control the level of correlation between the graphs.
This model is equivalent to $ER(n,\bfp)$ with 
\begin{IEEEeqnarray*}{rCl}
  p_{11} &=& rs_as_b \\
  p_{10} &=& rs_a(1-s_b) \\
  p_{01} &=& r(1-s_a)s_b \\
  p_{00} &=& 1-r(s_a + s_b - s_as_b).
\end{IEEEeqnarray*}
Solving for $r$ from the above definitions, we obtain
\begin{equation}
  r = \frac{(p_{10}+p_{11})(p_{01}+p_{11})}{p_{11}} = p_{11} + p_{10} + p_{01} + \frac{p_{10}p_{01}}{p_{11}}.
  \label{r}
\end{equation}
Observe that when $G_a$ and $G_b$ are independent, we have $r = 1$.
This reveals that the subsampling model is capable of representing any correlated \ER distribution with nonnegative correlation.
From \eqref{r}, we see that $r = \mathcal{O} \left(\frac{1}{\log n}\right)$ is equivalent to $p_{00} = 1 - \mathcal{O} \left(\frac{1}{\log n}\right)$ and $\frac{p_{01}p_{10}}{p_{00}p_{11}}  = \mathcal{O}\left(\frac{1}{\log n}\right)$.

\section{Graphs and cyclic sequences}
\label{section:cyclic-seq}
Let $w$ be a matrix of formal variables indexed by $[2] \times [2]$:
\[
  w = \twobytwo{w_{00}}{w_{01}}{w_{10}}{w_{11}}
\]
and let $z$ be a single formal variable.
For a set $\mathcal{S}$ and a permutation $\tau : \mathcal{S} \to \mathcal{S}$, define the generating function
\begin{align*}
  A_{\mathcal{S},\tau}(w,z) &= \sum_{g \in [2]^\mathcal{S}} \sum_{h \in [2]^\mathcal{S}} z^{\delta(\tau;g,h)} w^{\mu(g,h)}
\end{align*}

When $\mathcal{S} = \binom{[n]}{2}$ and $(G_a,G_b) \sim ER(\bfp)$, $A_{\mathcal{S},\tau}$ captures the joint distribution of $\mu(G_a,G_b)$ and $\delta(\tau;G_a,G_b)$:
\begin{equation*}
  P[\mu(G_a,G_b) = k, \delta(\tau; G_a,G_b) = i]
  = [w^k z^i]  A_{\mathcal{S}, \tau}\left(\bfp \odot w, z\right)
\end{equation*}
where $\bfp \odot w$ is the element-wise product of the matrices $\bfp$ and $w$.
This follows immediately from the definition of the $ER(\bfp)$ distribution.
\TODO{example}
\subsection{Generating functions}
\begin{definition}
  Let $\mathcal{S}$ be an finite index set and let $\sigma : \mathcal{S} \to \mathcal{S}$ be a permutation consisting of a single cycle of length $|\mathcal{S}|$.
  A cyclic $\mathcal{T}$-ary sequence is a pair $(\sigma, f)$ where $f : \mathcal{S} \to \mathcal{T}$.
\end{definition}
\bcomment{
  \begin{example}
    Let $\mathcal{S} = [3]$, $T = [2]$, $\sigma = x \mapsto (x+1) \bmod 3$, $f(0) = 0$, $f(1) = 0$, and $f(2) = 1$.
  \end{example}}
Let $\sigma$ be a permutation of $[\ell]$ with a single cycle.
For any such choices of $\sigma$, the sets of cyclic sequences obtained are isomorphic, so we can define the generating function
\[
  a_\ell(w,z) = A_{[\ell],\sigma}(w,z).
\]

\begin{lemma}
  \label{lemma:cycle-decomp}
  Let $\tau : \mathcal{S} \to \mathcal{S}$ be a permutation.
  Let $t_\ell$ be the number of cycles of length $\ell$ in $\tau$.
  Then $\sum_\ell \ell t_\ell = |\mathcal{S}|$ and
  \[    
    A_{\mathcal{S},\tau}(w,z) = \prod_{\ell \in \N} a_\ell(w,z)^{t_\ell} .
  \]
\end{lemma}
\begin{IEEEproof}
  Let $\gamma(a,b,c) = \frac{1}{2}(\one\{a \neq c\} - \one\{b \neq c\})$, so
  \[
    \delta(\tau;g,h) = \sum_{e \in \mathcal{S}} \gamma(g(e),g(\tau(e)),h(e)).
  \]
  Let $\mathcal{T}$ be the partition of $\mathcal{S}$ from the cycle decomposition of $\tau$.
  Then we have an alternate expression for $A_{\mathcal{S},\tau}$:
  \begin{align*}
    &\eq A_{\mathcal{S},\tau}(w,z)\\
    &= \sum_{g \in [2]^\mathcal{S}} \sum_{h \in [2]^\mathcal{S}} \prod_{e \in \mathcal{S}} z^{\gamma(g(e),g(\tau(e)),h(e))} w_{g(e),h(e)}\\
    &= \sum_{g \in [2]^\mathcal{S}} \sum_{h \in [2]^\mathcal{S}} \prod_{\mathcal{S}_i \in \mathcal{T}} \prod_{e \in \mathcal{S}_i} z^{\gamma(g(e),g(\tau(e)),h(e))} w_{g(e),h(e)}\\
    &\eql{a} \prod_{\mathcal{S}_i \in \mathcal{T}} \sum_{g \in [2]^{\mathcal{S}_i}} \sum_{h \in [2]^{\mathcal{S}_i}} \prod_{e \in \mathcal{S}_i} z^{\gamma(g(e),g(\tau(e)),h(e))} w_{g(e),h(e)}\\
    &= \prod_{\mathcal{S}_i \in \mathcal{T}} a_{|\mathcal{S}_i|}(w,z).
  \end{align*}
  In $(a)$, we use the fact that $e \in \mathcal{S}_i$ implies $\tau(e) \in \mathcal{S}_i$.
\end{IEEEproof}

For $l=1$, the generating function $a_1(x,y)$ is very simple.
There are 4 possible pairs of cyclic $[2]$-ary sequences of length one.
A cycle of length one in a permutation is a fixed point, so these cyclic sequences are unchanged by the application of $\sigma$ and $\delta(\sigma;g,h)$ is zero for each of them.
Thus $a_1(w,z) = (w_{00}+w_{01}+w_{10}+w_{11})$.

We define
\[
  \Atl_{\mathcal{S},\tau}(w,z) = \prod_{\ell \geq 2} a_{\ell}(w,z)^{t_{\ell}}.
\]
Just as $A_{\mathcal{S},\tau}$ captures the joint distribution of $M$ and $\delta(\tau)$, $\Atl_{\mathcal{S},\tau}$ captures the joint distribution of $\Mtl$ and $\delta(\tau)$.
Because $z$ does not appear in $a_1(w,z)$, we have 
\[
  [z^i] A_{\mathcal{S},\tau}(w,z) = a_1(w,z)^{t_1} [z^i] \Atl_{\mathcal{S},\tau}(w,z).
\]
This implies that $\delta(\tau;G_a,G_b)$ and $M$ are conditionally independent given $\Mtl$.
\subsection{Nontrivial cycles}
For $l=2$, there are 16 possible pairs of sequences.
There are only 4 pairs for which $\delta(\sigma;g,h) \neq 0$: the cases where $g$ and $h$ are each either $(0,1)$ or $(1,0)$.
In the two cases where $g = h$, 
$\mu(g,h) = \left(
  \begin{array}{rr} 1 & 0 \\ 0 & 1 
  \end{array}
\right)$,
$\mu(g \circ \sigma,h) = \left(
  \begin{array}{rr} 0 & 1 \\ 1 & 0 
  \end{array}
\right)$,
and $\delta(\sigma;g,h) = 1$.
In the two cases where $g \neq h$, $\delta(\sigma;g,h) = -1$.
Thus
\begin{multline}
  a_2(w,z) = (w_{00}+w_{01}+w_{10}+w_{11})^2\\ + 2w_{00}w_{11}(z - 1) + 2w_{01}w_{10}(z^{-1} - 1). \label{a-two}
\end{multline}

The following theorem relates longer cycles to cycles of length two.
\newcommand{\thmgf}{%
  Let $w \in \R_{> 0}^{[2] \times [2]}$, and $z \in \R$.
  Then for all $\ell \geq 2$, $a_{\ell}(w,z) \leq a_2(w,z)^{\ell/2}$.}

\begin{theorem}
  \label{thm:l-cycle-two-cycle}
  \thmgf
\end{theorem}
The proofs of Theorem~\ref{thm:l-cycle-two-cycle} and several intermediate lemmas are in Appendix~\ref{app:gf}.

\subsection{Tail bounds from generating functions}
\label{subsection:tail}

The following lemma is a well known inequality that we will apply in the proof of Theorem~\ref{thm:opt-z} and in several other places.
\begin{lemma}
  \label{lemma:chernoff}
  For a generating function $g(z) = \sum_i g_i z^i$
  where $g_i \geq 0$ for all $i \in \Z$ and a real number $0 \leq z_1 \leq 1$, 
  \[
    \sum_{i \leq j} [z^i] g(z) \leq z_1^{-j} g(z_1).
  \]
\end{lemma}
\begin{IEEEproof}
  \begin{equation*}
    \sum_{i \leq j} [z^i] g(z) 
    = \sum_{i \leq j} g_i
    \leq \sum_i g_i z_1^{i-j} 
    = z_1^{-j} g(z_1).\IEEEQEDhereeqn
  \end{equation*}
\end{IEEEproof}

\begin{theorem}
  \label{thm:opt-z}
  For $w \in \R_{> 0}^{[2] \times [2]}$ such that $w_{01}w_{10} < w_{00}w_{11}$, we have
  \begin{multline*}
    \sum_{i \leq 0} [z^i] \Atl_{\mathcal{S},\tau}(w,z) \leq \big((w_{00}+w_{01}+w_{10}+w_{11})^2\\
      - 2\left( \sqrt{w_{00}w_{11}} - \sqrt{w_{01}w_{10}} \right)^2 \big)^{\ttl/2} .
  \end{multline*}
\end{theorem}
\begin{IEEEproof}
  For all $0 \leq z_1 \leq 1$, we have
  \begin{align}
    \sum_{i \leq 0} [z^i] \Atl_{\mathcal{S},\tau}(w,z)
    &\eql{a} \sum_{i \leq 0} [z^i] \prod_{\ell \geq 2} a_\ell(w,z)^{t_\ell} \nonumber\\
    &\leql{b} \prod_{\ell \geq 2} a_{\ell}(w,z_1)^{t_\ell} \nonumber\\
    &\leql{c} a_2(w,z_1)^{\ttl/2} \label{a-two}
  \end{align}
  where $(a)$ follows from Lemma~\ref{lemma:cycle-decomp}, 
  $(b)$ follows from by Lemma~\ref{lemma:chernoff},
  and $(c)$ follows from Theorem~\ref{thm:l-cycle-two-cycle}
  and the definition $\ttl = \sum_{l=2}^n \ell t_{\ell}$.

  From \eqref{a-two}, $a_2(w,z) = u^2+2v$ where
  \begin{align*}
    u &= w_{00}+w_{01}+w_{10}+w_{11}\\
    v &= w_{00}w_{11} (z - 1) + w_{01}w_{10} (z^{-1} - 1).
  \end{align*}
  We would like to choose $z$ to minimize $v$.
  Over all positive $z$, the optimal choice is $z_1 = \left(\frac{w_{01}w_{10}}{w_{00}w_{11}}\right)^{1/2}$.
  We have $z_1 < 1$, so this can be used in \eqref{a-two}.
  Substituting $z_1$ into the expression for $v$, we obtain
  \begin{align}
    &\eq \min_{z} w_{00}w_{11} z - w_{00}w_{11} - w_{01}w_{10} + w_{01}w_{10} z^{-1}\nonumber\\
    &=   2\sqrt{w_{00}w_{01}w_{10}w_{11} z_1} - w_{00}w_{11} - w_{01}w_{10}\nonumber\\ 
    &=  -(\sqrt{w_{00}w_{11}}  - \sqrt{w_{01}w_{10}} )^2 \label{min-z}.
  \end{align}
  Combining this with $a_2(w,z) = u^2+2v$ and \eqref{a-two} gives the claimed bound.
\end{IEEEproof}

\subsection{Hypergeometric and binomial g.f.}
Chv\'{a}tal provided an upper bound on the tail probabilities of a hypergeometric random variable \cite{chvatal_tail_1979}.
The following lemma is essentially a translation of that bound into the language of generating functions.
\begin{lemma}
  \label{lemma:hyper}
  For all $a,b,n \in \N$, $a \leq n$ and $b \leq n$, and all $z \in \R$, $z > 0$
  \[
    \text{Hyp}(a,b,n;z) \leq \text{Bin}(a,b,n;z)\,.
  \]
\end{lemma}
\begin{IEEEproof}
  First, we have
  \begin{align*}
    &\binom{n}{a}\binom{n}{b} \text{Hyp}(a,b,n;z)\\
    &= [x^a y^b] (1 + x + y + xyz)^n\\
    &= [x^a y^b] ((1 + x)(1 + y) + xy(z-1))^n\\
    &= \sum_{\ell} \binom{n}{\ell} [x^a y^b] ((1 + x)(1 + y))^{n-\ell}(xy(z-1))^{\ell}\\
    &= \sum_{\ell} \binom{n}{\ell} [x^{a-\ell}] (1 + x)^{n-\ell} [y^{b-\ell}] (1 + y)^{n-\ell}(z-1)^{\ell}\\
    &= \sum_{\ell} \binom{n}{\ell} \binom{n-\ell}{a-\ell} \binom{n-\ell}{b-\ell} (z-1)^{\ell}\\
    &= \binom{n}{a} \binom{n}{b} \sum_{\ell} \frac{\binom{a}{\ell} \binom{b}{\ell}}{\binom{n}{\ell}} (z-1)^{\ell} .
  \end{align*}
  Observe that 
  \[
    \frac{\binom{b}{\ell}}{\binom{n}{\ell}}\frac{n^{\ell}}{b^{\ell}}
    = \prod_{i=0}^{\ell-1}\frac{(b-i)n}{(n-i)b}
    = \prod_{i=0}^{\ell-1}\frac{bn-in}{bn-ib}
    \leq \prod_{i=0}^{\ell-1} 1 = 1.
  \]
  Thus for $z \geq 1$,
  \begin{align*}
    \text{Hyp}(a,b,n;z)
    &\leq \sum_{\ell} \binom{a}{\ell} \frac{b^{\ell}}{n^{\ell}} (z-1)^{\ell} \\
    &= \left(1 + \frac{b}{n} (z-1)\right)^a \\
    &= \text{Bin}(a,b,n;z).
  \end{align*}
  
  If $0 < z \leq 1$, then $z^{-1} \geq 1$ and 
  \begin{align*}
    \text{Hyp}(a,b,n;z) 
    &= z^a \text{Hyp}(a,n-b,n;z^{-1})\\
    &\leq z^a \text{Bin}(a,n-b,n;z^{-1})\\
    &= \text{Bin}(a,b,n;z),
  \end{align*}
  which completes the proof.
\end{IEEEproof}


\section{Achievability theorems}
In this section we will prove Theorem~\ref{thm:main}.
This is the combination of two results for different values of $p_{11}$.
Corollary~\ref{cor:sparse} in Section~\ref{section:ach-one} handles
$\frac{2 \log n + \omega(1)}{n} \leq p_{11}$ and Theorem~\ref{thm:ach-two} in Section~\ref{section:ach-two} handles $\frac{\log n + \omega(1)}{n} \leq p_{11} \leq \mathcal{O}\left(\frac{\log n}{n}\right)$.

\subsection{Permutations}
\label{section:permutations}
Let $S_{n,\ntl}$ be the set of permutations of $[n]$ with exactly $n-\ntl$ fixed points.
If $\pi \in S_{n,\ntl}$, then $e=\{i,j\}$ is a fixed point of $\tau$ if either $i$ and $j$ are both fixed points of $\pi$ or $i$ and $j$ form a cycle of length 2 in $\pi$.
Thus $t_1$, the number of fixed points of $\tau$, satisfies
\begin{equation*}
  \binom{n-\ntl}{2} \leq t_1 \leq \binom{n-\ntl}{2} + \frac{\ntl}{2}. 
\end{equation*}
We will need a number of variations on these inequalities and we collect them in this section.
Let $\nu = \ntl/n$. Then
\begin{align}
  \frac{t_1}{t}
  &\leq \frac{(n - \ntl)(n-1-\ntl) + \ntl}{n(n-1)}\nonumber\\
  &= (1 - \nu)\left(1 - \frac{n}{n-1}\nu\right) + \frac{\nu}{n-1}\nonumber\\
  &= (1-\nu)^2 + \frac{\nu}{n-1}(1 - (1-\nu))\nonumber\\
  &= (1-\nu)^2 + \frac{\nu^2}{n-1}.\label{nu-bound}
\end{align}
For $0 \leq \nu \leq 1$, the best linear upper bound on $t_1$ is
\begin{equation*}
  \frac{t_1}{t} \leq 1 - \frac{\nu(n-2)}{n-1}
\end{equation*}
Equivalently,
\begin{equation}
\ttl = t - t_1 \geq \frac{\ntl(n-2)}{2} \label{linear-bound}.
\end{equation}
In the other direction, we have
\begin{align}
  \ttl
  &\leq \binom{n}{2} - \binom{n-\ntl}{2}\nonumber\\
  &= \frac{n\ntl + (n-1)\ntl - \ntl^2}{2}\nonumber\\
  &\leq n\ntl. \label{ttl-ub}
\end{align}

\bcomment{
Recall that $\ttl = t - t_1$, the number of vertex pairs in nontrivial cycles of $\tau$, so 
\begin{align}
  \ttl 
  &\geq \frac{n(n-1) - (n-\ntl)(n-\ntl-1)-\ntl}{2}\nonumber\\
  &= \frac{\ntl(2n-\ntl-2)}{2}. \label{c-bound}
\end{align}
}

\begin{lemma}
  \label{lemma:perm-gf}
  For $z \in \R$ such that $0 \leq z < n^{-1}$,
  \[
    \sum_{\ntl} |S_{n,\ntl}|z^{\ntl} \leq 1 + \frac{n^2z^2}{1-nz} .
  \]
\end{lemma}
\begin{IEEEproof}
  We have $|S_{n,\ntl}| = \binom{n}{\ntl} (!\ntl)$, where $!\ntl$ is the number of derangements of $[\ntl]$, i.e. the permutations of $[\ntl]$ with no fixed points.
  Note that $!0 = 1$ and $!1 = 0$, so $|S_{n,0}| = 1$ and $|S_{n,1}| = 0$.
  For $\ntl \geq 2$, we use $|S_{n,\ntl}| \leq \binom{n}{\ntl}\ntl! \leq n^{\ntl}$.
  For $0 \leq x < 1$, $\sum_{\ntl \geq 2}^n x^{\ntl} \leq \sum_{\ntl \geq 2} x^{\ntl} = \frac{x^2}{1-x}$ .
\end{IEEEproof}
We will apply Lemma~\ref{lemma:perm-gf} to compute union bounds over error events associated with permutations.
The upper bound on $z$ is slightly inconvenient, but it can be eliminated as follows.
\begin{lemma}
  \label{lemma:union-bound}
  Let $\{\eta_{\pi} : \pi \in S_n \setminus \text{id}\}$ be a family of events.
  If $P[\eta_{\pi}] \leq z^{\ntl}$ for all $\pi \in S_{n,\ntl}$, then $P[\vee_{\pi \neq \text{id}} \eta_\pi] \leq 3 n^2 z^2$ .
\end{lemma}

\begin{IEEEproof}
  If $nz \leq \frac{2}{3}$, the bound follows from the union bound and Lemma~\ref{lemma:perm-gf}:
  \bcomment{
    \begin{align*}
      P [ \vee_{\pi \neq \text{id}} \eta_{\pi}]
      &\leq \sum_{\pi \neq id} P [\eta_{\pi}]\\
      &\leq \sum_{\ntl=2}^n |S_{n,\ntl}| z^{\ntl}\\
      &\leq \frac{n^2z^2}{1-nz}\\
      &\leq 3n^2z^2.
    \end{align*}
  }
  \[
    P \bigg[ \bigvee_{\pi \neq \text{id}} \eta_{\pi}\bigg]
    \leq \sum_{\pi \neq \text{id}} P [\eta_{\pi}]
    \leq \sum_{\ntl=2}^n |S_{n,\ntl}| z^{\ntl}
    \leq \frac{n^2z^2}{1-nz}
    \leq 3n^2z^2.
    \]  
    If $nz \geq \frac{2}{3}$, then $P[\vee_{\pi \neq \text{id}} \eta_\pi] \leq 1 < \frac{4}{3} \leq 3n^2z^2 $.
\end{IEEEproof}

\subsection{Achievability theorem for dense graphs}
\label{section:ach-one}

We have developed enough tools to prove the first achievability theorem.
\begin{lemma}
  \label{lemma:dense-ub}
  Let $\bfp$ satisfy $p_{01}p_{10} < p_{11}p_{00}$.
  For all $\pi \in S_{n,\ntl}$,
  \[
    P [\delta(\tau) \leq 0 ] \leq z_2^{\ntl}
  \]
  where $\tau = l(\pi)$ and 
  \[
    z_2 = \exp\left(-\frac{1}{2}(n-2)(\sqrt{p_{11}p_{00}} - \sqrt{p_{01}p_{10}})^2\right) .
  \]  
\end{lemma}
\begin{IEEEproof}
  For all $\pi \in S_{n,\ntl}$,
  \begin{align*}
    P [\delta(\tau) \leq 0 ] 
    &= \sum_{i \leq 0} [z^i] \Atl_\tau(\bfp,z) \\
    &\leql{a} (1 - 2 (\sqrt{p_{11}p_{00}} - \sqrt{p_{01}p_{10}})^2)^{\ttl/2} \\
    &\leql{b} \exp(- \ttl (\sqrt{p_{11}p_{00}} - \sqrt{p_{01}p_{10}})^2)\\
    &\leql{c} \exp\left(-  \frac{\ntl(n-2)(\sqrt{p_{11}p_{00}} - \sqrt{p_{01}p_{10}})^2}{2}\right)
  \end{align*}
  where $(a)$ follows from Theorem~\ref{thm:opt-z} with $w = \bfp$, $(b)$ uses $1 + x \leq e^x$, and $(c)$ uses \eqref{linear-bound}.
\end{IEEEproof}

\newtheorem*{T2}{Theorem~\ref{thm:ach-one}}
\begin{T2}
  \thmachone
\end{T2}
\begin{IEEEproof}
  From Lemma~\ref{lemma:dense-ub}, for all $\pi \in S_{n,\ntl}$, $P [\delta(\tau) \leq 0 ] \leq z_2^{\ntl}$ 
  Applying Lemma~\ref{lemma:union-bound},
  \begin{equation*}
    P [ \vee_{\pi \neq id} \delta(\tau) \leq 0 ]
    \leq 3n^2z_2^2
  \end{equation*}
  which is $o(1)$ whenever $nz_2$ is $o(1)$. 
\end{IEEEproof}

Specializing Theorem~\ref{thm:ach-one} it to the conditions of Theorem~\ref{thm:main} gives the following.
\begin{corollary}
  \label{cor:sparse}
  Let $\bfp$ satisfy the conditions \eqref{p11-sparse}, \eqref{p-sparse}, and \eqref{p-corr}  and let 
  \[
    \frac{2 \log n + \omega(1)}{n} \leq p_{11} .
  \]
  Then the MAP estimator is correct with probability $1 - o(1)$.
\end{corollary}
\begin{IEEEproof}
  In this regime, both $p_{00} = 1 - p_{01} - p_{10} - p_{11}$ and $1 - \sqrt{\frac{p_{01}p_{10}}{p_{00}p_{11}}}$ are $1 - \mathcal{O} \left( \frac{1}{\log n} \right)$, so the condition of Theorem~\ref{thm:ach-one} is equivalent to the claimed condition.
\end{IEEEproof}
Note that \eqref{p-corr} is slightly stronger than what we require.

\subsection{Achievability theorem for sparse graphs}
Throughout this section, we will replace \eqref{p11-sparse} with a much strong sparsity constraint on $G_a \wedge G_b$:
\begin{align}
  p_{11} &\leq \mathcal{O}\left(\frac{\log n}{n}\right). \label{p11-v-sparse}
\end{align}

\label{section:ach-two}

\begin{lemma}
  \label{lemma:sparse}
  Let $\bfp$ satisfy the conditions \eqref{p-sparse}, \eqref{p-corr}, and \eqref{p11-v-sparse} and let $\mtl \leq \mathcal{O}\left(\frac{\ttl \log n}{n}\right)$.
  Then for all $\ntl$ and all $\pi \in S_{n,\ntl}$,
  \[
    P[\delta(\tau) \leq 0 | \Mtl = \mtl] \leq z_4^{\mtl} z_5^{\ntl},
  \]
  where $z_4 = \mathcal{O} \left(\frac{1}{\log n}\right)$ and $z_5 = \mathcal{O}(1)$.
\end{lemma}
We have $\Mtl \sim \text{Bin}(\ttl,p_{11},1)$, so $\E[\Mtl] = \ttl p_{11} \leq \mathcal{O}\left(\frac{\ttl \log n}{n}\right)$.
  Thus the assumption $\mtl \leq \mathcal{O}\left(\frac{\ttl \log n}{n}\right)$ excludes the possibility that $\Mtl$ is more than a constant factor larger than its expected value.
  In this regime, each additional appearance of $(1,1)$ in the nontrivial region significantly reduces the probability that $\delta(\tau) \leq 0$.
  If $\mtl$ is large enough some 2-cycles contain two appearances of $(1,1)$.
  These collisions contribute nothing to $\delta(\tau)$ and diminish the return we receive for increasing $\mtl$.
  Thus the condition $\mtl \leq \mathcal{O}\left(\frac{\ttl \log n}{n}\right)$ is required.

  Similarly, condition \eqref{p-sparse}, $p_{01} +p_{10} \leq \mathcal{O}{((\log n)^{-1})}$, is required to ensure that the number of 2-cycles containing a $(1,1)$ and a $(1,0)$ is negligible. 
  \begin{IEEEproof}
    For any $w_* \geq 0$ such that $w_*p_{11}p_{00} \geq p_{01}p_{10}$, we have
  \begin{align}
    &P[\delta(\tau) \leq 0, \Mtl = \mtl]\nonumber\\
    &= \sum_{d \leq 0} [z^d w_{11}^{\mtl}] \Atl_{\mathcal{S}, \tau}\left(\twobytwo{p_{00}}{p_{01}}{p_{10}}{p_{11}w_{11}}, z\right)\nonumber\\
    &\leql{a} \sum_{d \leq 0} [z^d] w_*^{-\mtl} \Atl_{\mathcal{S}, \tau}\left(\twobytwo{p_{00}}{p_{01}}{p_{10}}{p_{11}w_*}, z\right)\nonumber\\
    &\leql{b} w_*^{-\mtl} \alpha(\bfp,w_*)^{\ttl/2} \label{p-tau-mtl}
  \end{align}
  where $(a)$ follows from Lemma~\ref{lemma:chernoff}, $(b)$ follows from Theorem~\ref{thm:opt-z}, and $\alpha(\bfp,w_*)$ is 
  \[
    (1-p_{11}+p_{11}w_*)^2 - 2\left( \sqrt{p_{00}p_{11}w_*} - \sqrt{p_{01}p_{10}} \right)^2.
  \]
  We have
  \begin{align}
      P[\Mtl = \mtl]
      &= [w_{11}^{\mtl}] \Atl_{\mathcal{S}, \tau}(p \odot w, 1)\nonumber\\
      &= \binom{\ttl}{\mtl} p_{11}^{\mtl} (1-p_{11})^{\ttl - \mtl}\nonumber\\
      &\geq \left(\frac{\ttl}{\mtl}\right)^{\mtl} p_{11}^{\mtl} (1-p_{11})^{\ttl - \mtl}\nonumber\\
      &= \left(\frac{\ttl p_{11}}{\mtl (1-p_{11})}\right)^{\mtl}(1-p_{11})^{\ttl}.\label{p-mtl}
  \end{align}
  Let $p'_{ij} = p_{ij}/(1-p_{11})$.
  Combining \eqref{p-tau-mtl} and \eqref{p-mtl}, we get
  \begin{equation}
    P[\delta(\tau) \leq 0 | \Mtl = \mtl]
    \leq \left(\frac{\mtl}{\ttl p'_{11}w_*}\right)^{\mtl}\left(\frac{\alpha(\bfp,w_*)}{(1-p_{11})^2}\right)^{\ttl/2} . \label{p-tau-given-mtl}
  \end{equation}
  Now let
  \begin{equation*}
    w_* = \frac{1}{p'_{11}}\cdot\frac{\mtl \log n + \E[\Mtl]}{\ttl} = \frac{1}{ p'_{11}}\left(\frac{\mtl \log n}{\ttl} + p_{11}\right).
  \end{equation*}
  Because $w_* \geq \frac{p_{11}}{p'_{11}} = 1-p_{11}$ and $\frac{w_*p_{11}p_{00}}{p_{01}p_{10}} \geq \Omega((\log n)^3)$, $w_*$ satisfies the required condition for all sufficiently large $n$.
  With this choice of $w_*$, the first term of the right side of \eqref{p-tau-given-mtl} becomes
  \begin{equation*}
    \left(\frac{\mtl}{\ttl p'_{11}w_*}\right)^{\mtl} = \left(\frac{\mtl}{\mtl \log n + \ttl p_{11}}\right)^{\mtl} \leq \left(\frac{1}{\log n}\right)^{\mtl}.
  \end{equation*}
  The second term is 
  \begin{align*}
    &\eq \alpha(\bfp,w_*)(1-p_{11})^{-2}\\
    &=(1+p'_{11}w_*)^2 - 2\left( \sqrt{p'_{00}p'_{11}w_*} - \sqrt{p'_{01}p'_{10}} \right)^2\\
    &= 1 + (p'_{11}w_*)^2 + \\
    &\eq 2 p'_{11}w_* - 2p'_{00}p'_{11}w_* - 2p'_{01}p'_{10} + 4\sqrt{p'_{00}p'_{01}p'_{10}p'_{11}w_*} \\
    &\leq 1 + (p'_{11}w_*)^2 + 2 (p'_{01}+p'_{10})p'_{11}w_* + 4\sqrt{p'_{00}p'_{01}p'_{10}p'_{11}w_*}
  \end{align*}
  where we used $1 - p'_{00} = p'_{01}+p'_{10}$.
  Applying $\log(1+x) \leq x$, we obtain
  \begin{align*}
    &\eq \frac{\ttl}{2} \log \left(\alpha(\bfp,w_*)(1-p_{11})^{-2}\right)\\
    &\leq \frac{\ttl}{2}(p'_{11}w_*)^2 + \ttl (p'_{01}+p'_{10})p'_{11}w_* + 2 \ttl\sqrt{p'_{00}p'_{01}p'_{10}p'_{11}w_*}.
  \end{align*}
  We can use \eqref{ttl-ub} ($\ttl \leq n \ntl$) and $\ttl^{-1} \leq \mathcal{O}\left(\frac{\log n}{n \mtl}\right)$ to eliminate appearances of $\ttl$ and use \eqref{p11-v-sparse} to eliminate $p_{11}$.
  We have
  \begin{align*}
    \ttl(p'_{11}w_*)^2
    &= \frac{\mtl^2 (\log n)^2}{\ttl} + 2 \mtl (\log n) p_{11} + \ttl p_{11}^2\\
    & \leq \mathcal{O}\left( \frac{\mtl(\log n)^3}{n} + \frac{\mtl(\log n)^2}{n} + \frac{\ntl(\log n)^2}{n}\right)\\
    &\leq \mathcal{O}(\mtl + \ntl).
  \end{align*}
  Condition \eqref{p-sparse} allows us to bound $p_{01} + p_{10}$:
  \begin{align*}
    \ttl (p'_{01}+p'_{10})p'_{11}w_*
    &= (p'_{01}+p'_{10})(\mtl (\log n) + \ttl p_{11})\\
    &\leq \mathcal{O}\left((\log n)^{-1}(\mtl (\log n) + \ntl (\log n))\right)\\
    &\leq \mathcal{O}(\mtl + \ntl).
  \end{align*}
  Condition \eqref{p-corr} allows us to bound $p_{01}p_{10}$:
  \begin{align*}
    &\eq \ttl\sqrt{p'_{00}p'_{01}p'_{10}p'_{11}w_*}\\
    &= \ttl\sqrt{p'_{00}p'_{01}p'_{10}(\mtl (\log n)\ttl^{-1} + p_{11})}\\
    &\leq \mathcal{O}\left( \ttl \sqrt{(\log n)^{-3}p_{11}((\log n)^2n^{-1} + p_{11})}\right)\\
    &\leq \mathcal{O}\left(\ttl n^{-1}\right)\\
    &\leq \mathcal{O}(\ntl).
  \end{align*}
  Overall, we have 
  \begin{align*}
    P[\delta(\tau) \leq 0 | \Mtl = \mtl] 
    &\leq \left(\frac{1}{\log n}\right)^{\mtl} (e^{\mathcal{O}(1)})^{\mtl} (e^{\mathcal{O}(1)})^{\ntl} \\
    &\leq \left(\mathcal{O}\left(\frac{1}{\log n}\right)\right)^{\mtl}\left(\mathcal{O}(1)\right)^{\ntl} . \IEEEQEDhere
  \end{align*}
\end{IEEEproof}

\begin{lemma}
  \label{lemma:m-ub}
  Let $\bfp$ satisfy the conditions \eqref{p-sparse}, \eqref{p-corr}, and \eqref{p11-v-sparse} and let $m \leq \mathcal{O}(n \log n)$.
  Then for all $\ntl$ and all $\pi \in S_{n,\ntl}$,
  \[
    P[ \delta(\tau) \leq 0 | M = m] \leq z_7^{\ntl} .
  \]
  where 
  $z_7 \leq \exp\left(- \frac{2m}{n} + \mathcal{O}(1)\right)$.
\end{lemma}
\begin{IEEEproof}
  For a fixed $\pi$, and thus a fixed $\tau$, we condition on $\Mtl$, the number of edges of $G_a \wedge G_b$ that are in nontrivial cycles of $\tau$.
  We do this because we need some upper bound on $\mtl$ to apply Lemma~\ref{lemma:sparse}.
  Recall that $\Mtl | M = m$ has a hypergeometric distribution and that $\E[\Mtl | M = m] = \frac{m \ttl}{t}$.
  Define $\mtl^* = e^2 \frac{m \ttl}{t}$ and write
  \begin{align*}
      P [\delta(\tau) \leq 0 | M = m]
     &= P[\delta(\tau) \leq 0, \Mtl \leq \mtl^* | M = m]\\
     &\eq + P[\delta(\tau) \leq 0, \Mtl > \mtl^* | M = m].
  \end{align*}

  The first error term contains the typical values of $\Mtl$.
  \begin{align*}
    \epsilon_1
    &\deq P[\delta(\tau) \leq 0, \Mtl \leq \mtl^* | M = m]\\
    &\eql{a}  \sum_{\mtl \leq \mtl^*} P[\Mtl = \mtl | M = m] P [\delta(\tau) \leq 0 |\Mtl = \mtl] \\
    &\leql{b} \sum_{\mtl \leq \mtl^*} P[\Mtl = \mtl | M = m] z_4^{\mtl} z_5^{\ntl}\\
    &\leq     z_5^{\ntl} \sum_{\mtl} P[\Mtl = \mtl | M = m] z_4^{\mtl}\\
    &\eql{c}  z_5^{\ntl} \text{Hyp}(m, \ttl, t; z_4)\\
    &\leql{d} z_5^{\ntl} \text{Bin}(m, \ttl, t; z_4)\\
    &= z_5^{\ntl} \left(1 + \frac{\ttl}{t}(z_4-1)\right)^m\\
    &= z_5^{\ntl} \left(1 + \left(\frac{t_1}{t} - 1\right)(1 - z_4)\right)^m
  \end{align*}
  where $(a)$ uses the conditionally independence of $\delta(\tau)$ and $\Mtl$ given $M$, $(b)$ follows from Lemma~\ref{lemma:sparse},
  $(c)$ follows from \eqref{mtl-gf-hyper}, and $(d)$ follows from Lemma~\ref{lemma:hyper}.

  Let $\nu = \frac{\ntl}{n}$.
  For sufficiently large $n$, we have $z_4 < 1$ so we can apply \eqref{nu-bound}:
  \begin{align*}
    &\eq 1 + \left(\frac{t_1}{t} - 1\right)(1 - z_4)\\
    &\leq 1 + \left((1-\nu)^2 + \frac{\nu^2}{n-1} - 1\right)(1 - z_4)\\
    &= (1-\nu)^2 + \frac{\nu^2}{n-1}(1 - z^4) +\nu(2-\nu)z_4\\
    &\leq (1-\nu)^2 + \frac{\nu^2}{n-1} +2 \nu z_4.
  \end{align*}
  Now let $z_6 = (n-1)^{-1} + 2z_4$.
  We will handle small and large values of $\ntl$ separately.
  In the region $2 \leq \ntl \leq n(1 - e^{-1})$, or equivalently $e^{-1} \leq 1-\nu \leq 1$, we have
  \begin{align*}
    &\eq  (1 - \nu)^2 + \nu z_6\\
    &=    (1 - \nu) \left( 1 - \nu\left(1 -  \frac{z_6}{1-\nu} \right)\right)\\
    &\leq (1 - \nu) (1 - \nu(1 - e z_6))\\
    &\leq e^{-\nu}\exp(-\nu(1 - e z_6))\\
    &=    \exp(-\nu(2 - ez_6))
  \end{align*}
  In the region $n(1 - e^{-1}) < \ntl \leq n$, or equivalently $0 \leq 1-\nu \leq e^{-1}$, we have
  \begin{align*}
    &\eq  (1 - \nu)^2 + \nu z_6\\
    &\leq e^{-1}(1 - \nu) + \nu z_6\\
    &=    e^{-1}(1 - \nu(1 - e z_6))\\
    &\leq e^{-\nu} \exp(- \nu(1 - e z_6))\\
    &=    \exp(-\nu(2 - ez_6 ))
  \end{align*}

  Thus
  \begin{align*}
    \log \epsilon_1
    &\leq \ntl \log z_5 - \frac{m\ntl}{n}\left(2-ez_6 \right) \\
    &= \ntl \left(- \frac{2m}{n} + \frac{e m z_6}{n} + \log z_5\right)\\
    &\leq \ntl \left(- \frac{2m}{n} + \mathcal{O}(1)\right)
  \end{align*}
  because $z_6 = 2 z_4 + \frac{1}{n-1}$, $z_4 = \mathcal{O} \left( \frac{1}{\log n}\right)$, $\frac{m}{n} = \mathcal{O}(\log n)$,
  and $\log z_5 = \mathcal{O}(1)$.

  The second error term is small compared to the first, so we do not need to obtain the best possible exponent. We have
  \begin{align*}
    \epsilon_2
    &\deq P[\delta(\tau) \leq 0, \Mtl > \mtl^* | M = m]\\
    &\leq P[\Mtl > \mtl^* | M = m]\\
    &\leql{a} \text{Hyp}(m, \ttl, t; z_3)z_3^{-e^2m\ttl/t}\\
    &\leql{b} \text{Bin}(m, \ttl, t; z_3)z_3^{-e^2m\ttl/t}\\
    &= \left(1 + \frac{\ttl}{t}(z_3-1)\right)^m z_3^{-e^2m\ttl/t}\\
    &\leq e^{(z_3 - 1)m \ttl/t} z_3^{-e^2m\ttl/t}\\
    &\eql{c} \exp\left((e^2 - 1)\frac{m\ttl}{t} - \frac{2 e^2 m\ttl}{t} \right)\\
    &=  \exp\left(-(e^2 + 1)\frac{m\ttl}{t} \right)\\
    &\leql{d} \exp\left(-(e^2 + 1)\frac{m \ntl (n-2)}{n(n-1)} \right)
  \end{align*}
  where $(a)$ follows from Lemma~\ref{lemma:chernoff},
  $(b)$ follows from Lemma~\ref{lemma:hyper},
  $(c)$ follows from letting $z_3 = e^2$,
  and $(d)$ follows from \eqref{linear-bound}.

  Thus $\epsilon_2$ is exponentially smaller than $\epsilon_1$ and we have the claimed bound.
\end{IEEEproof}

\begin{lemma}
  \label{lemma:m-union-bound}
  Let $p$ satisfy the conditions \eqref{p-sparse}, \eqref{p-corr}, and \eqref{p11-v-sparse} and let $m \leq \mathcal{O}(n \log n)$.
  Then
  \[
    P [ \vee_{\pi \neq id} \delta(\tau) \leq 0 | M = m] \leq \mathcal{O}(n^2 z_8^m)
  \]
  where
  $z_8 = \frac{n}{n+4}$.
\end{lemma}
\begin{IEEEproof}
  From Lemma~\ref{lemma:m-ub}, for all $\ntl$ and all $\pi \in S_{n,\ntl}$,
  \[
    P[ \delta(\tau) \leq 0 | M = m] \leq z_7^{\ntl}
  \]
  so by Lemma~\ref{lemma:union-bound},
  \begin{equation*}
    P [ \vee_{\pi \neq id} \delta(\tau) \leq 0 | M = m]
    \leq 3n^2z_7^2 \leq \mathcal{O}(n^2)\exp\left(- \frac{4m}{n}\right).
  \end{equation*}
  Finally
  \begin{equation*}
    \exp\left(-\frac{4}{n}\right) = \frac{1}{\exp\left(\frac{4}{n}\right)} \leq \frac{1}{1+\frac{4}{n}} = \frac{n}{n+4} = z_8.\IEEEQEDhereeqn
  \end{equation*}
\end{IEEEproof}

\begin{lemma}
  \label{lemma:m-averaging}\TODO{Add appropriate conditions about the probability space.}
  Let $\mcE$ be an event. If
  \begin{equation}
    P [ \mcE | M = m] \leq z_9 z_8^m \label{abstract-m-ub}
  \end{equation}
  for all $m \leq (1+\Omega(1))\E[M]$ and
  \begin{equation}
    p_{11} \geq \frac{\log z_9 + \omega(1)}{t(1-z_8)}, \label{abstract-p-lb}
  \end{equation}
  then $P[\mcE] \leq o(1)$.
\end{lemma}
\begin{IEEEproof}
  We have $M \sim \text{Bin}(t,p_{11},1)$, so $\E[M] = tp_{11} \geq n \log n$.
  Thus the probability that $M \geq (1+\epsilon)(tp_{11})$ is $o(1)$ for any $\epsilon > 0$.
  We have 
  \begin{equation*}
    P [ \mcE ]    \leq o(1) + P [ \mcE | M \leq (1+\epsilon)tp_{11}].
  \end{equation*}
  Now we analyze the main term:
  \begin{align*}
    &\eq  P [ \mcE  | M \leq (1+\epsilon)tp_{11}]\\
    &=    \sum_{m \leq (1+\epsilon)\E[M]} P [ \mcE | M = m] P[M = m] \\
    &\leql{a} z_9 \sum_{m \leq (1+\epsilon)\E[M]} z_8^m P[M = m] \\
    &\leq z_9 \sum_{m} z_8^m P[M = m] \\
    &= z_9 (1 + p_{11}(z_8-1))^t\\
    &\leql{b} \exp\left( \log z_9 + t p_{11}(z_8-1) \right)\\
    &\leql{c} \exp(-\omega(1))
  \end{align*}
  The inequality $(a)$ follows from \eqref{abstract-m-ub},
  $(b)$ follows from $1+x \leq e^x$, and 
  $(c)$ follows from \eqref{abstract-p-lb}.
\end{IEEEproof}
\newcommand*{\thmachtwo}{%
  Let $\bfp$ satisfy the conditions  \eqref{p11-main}, \eqref{p-sparse}, \eqref{p-corr}, and \eqref{p11-v-sparse}.
  Then the MAP estimator is correct with probability $1 - o(1)$.
}

\begin{theorem}
  \label{thm:ach-two}
  \thmachtwo
\end{theorem}
\bcomment{
\newtheorem*{T3}{Theorem~\ref{thm:ach-two}}
\begin{T3}
  \thmachtwo
\end{T3}}
\begin{IEEEproof}
  Let $\mcE = \vee_{\pi \neq id} (\delta(\tau) \leq 0)$.
  From Lemma~\ref{lemma:m-union-bound}, we have $P[\mcE|M=m] \leq z_9z_8^m$ with $z_9 = Cn^2$ for some constant $C$ and $z_8 = \frac{n}{n+4}$.
  From \eqref{p11-main}, we have
  \begin{align*}
    p_{11}
    &\geq \frac{\log n + \omega(1)}{n}\\
    &\geq \frac{\log n + \frac{5 \log n}{n-1} + \omega(1)\left(1+\frac{5}{n-1}\right)}{n}\\
    &\geq \frac{\log n + \omega(1)}{n}\left(1+\frac{5}{n-1}\right)\\
    &\geq \frac{\log n + \frac{1}{2} \log C + \omega(1)}{n}\left(\frac{n+4}{n-1}\right)\\
    &= \frac{\log (Cn^2) + \omega(1)}{\binom{n}{2}\frac{4}{n+4}}
  \end{align*}
  which is the condition \eqref{abstract-p-lb} that we require to apply Lemma~\ref{lemma:m-averaging}.
\end{IEEEproof}

\section{Proof of Converse}
\label{section:converse}
The converse statement depends on the following lemma.
\begin{lemma}
  \label{lemma:intersection}
  Let $g_a$ and $g_b$ be graphs on the vertex set $[n]$.
  For all $\pi \in Aut(g_a \wedge g_b)$, $\delta(\tau;g_a,g_b) \leq 0$.
\end{lemma}
\begin{IEEEproof}
  Let $\tau = l(\pi)$ and recall that
  \begin{IEEEeqnarray*}{rCl}
    \Delta(g_a,g_b) &=& \sum_{e \in \binom{[n]}{2}} |g_a(e) - g_b(e)|\\
    \delta(\tau;g_a,g_b) &=& \Delta(g_a \circ \tau; g_b) - \Delta(g_a, g_b).
  \end{IEEEeqnarray*}

  Let $e \in \binom{[n]}{2}$.
  Suppose that $(g_a,g_b)(e) = (1,1)$, so $(g_a \wedge g_b)(e) = 1$.
  Because $\pi \in Aut(g_a \wedge g_b)$, $(g_a \wedge g_b)(\tau(e)) = 1$.
  Then the contribution of $e$ 
  to both $\Delta(g_a,g_b)$ and $\Delta(g_a \circ \tau; g_b)$ is zero.

  Suppose $(g_a \wedge g_b)(e) = 0$.
  The cycle of $\tau$ containing $e$ is $S = \{\tau^i(e) : i \in \N\}$.
  For all $e' \in S$, $(g_a \wedge g_b)(e') = 0$ and $(g_a, g_b)(e')$ is $(0,0)$, $(0,1)$, or $(1,0)$.
  Thus the contribution of $S$ to $\Delta(g_a,g_b)$ is equal to total number of edges in $g_a$ and $g_b$ in $S$.
  The contribution of $S$ to $\Delta(g_a \circ \tau; g_b)$ cannot be larger.
\end{IEEEproof}

It is well-known that \ER graphs with average degree less than $\log n$ have many automorphisms \cite{bollobas_random_1998}.
The following lemma is precise version of this fact that is suitable for our purposes.
\TODO{low priority:no need for the two appearances of epsilon to be the same}
\begin{lemma}
  \label{lemma:isolated}
  Let $G \sim ER(n,p)$.
  If $p \leq \frac{\log n - \omega(1)}{n}$, then there is some sequence $\epsilon_n \to 0$ such that $P[|Aut(G)| \leq \epsilon_n^{-1}] \leq \epsilon_n$.
\end{lemma}
This follows easily from the second moment method.
Full details can be found in \cite{cullina_improved_2016}.
\bcomment{
\begin{IEEEproof}
  Let $X$ be the number of isolated vertices in $G$.
  A permutation that moves only isolated vertices is an automorphism of $G$, so $|Aut(G)| \geq X!$.
  We will use Chebyshev's inequality to bound the probability that there are few isolated vertices in $G$: 
  \begin{equation*}
    P[X \leq \frac{1}{2} \E[X]] \leq 4\frac{\E[X^2] - \E[X]^2}{\E[X]^2}.
  \end{equation*}

  The probability that a particular vertex is isolated is $(1-p)^{n-1}$.
  Thus $\E[X] = n (1-p)^{n-1}$.
  The probability that a particular pair of vertices are both isolated is $(1-p)^{2n-3}$.
  Thus $E\left[\binom{X}{2}\right] = \binom{n}{2} (1-p)^{2n-3}$.
  Then
  \begin{IEEEeqnarray*}{Cl}
    & \E[X^2]\E[X]^{-2} - 1\\
    =& \left(2 E\left[\binom{X}{2}\right] + \E[X]\right)\E[X]^{-2} - 1\\
    =& \frac{(n^2-n) (1-p)^{2n-3}}{n^2 (1-p)^{2n-2}} + E[X]^{-1} - 1\\
    =& (1-p)^{-1} - n^{-1}(1-p)^{-1} + E[X]^{-1} - 1\\
    \leq& p + E[X]^{-1}.
  \end{IEEEeqnarray*}
  Recall that $p \leq \frac{\log n}{n}$, so $p \to 0$.
  Finally we compute the limiting behavior of the expected value of $X$:
  \begin{IEEEeqnarray*}{rCl}
    \E[X] 
    &=& n(1-p)^{n-1}\\
    &=& n\left(1 + \frac{p}{1-p}\right)^{-(n-1)}\\
    &\geq& n\left(\exp\left(\frac{p}{1-p}\right)\right)^{-n}\\
    &=& \exp\left(\log n - \frac{np}{1-p}\right)\\
    &=& \exp\left(\frac{t_n - p \log n}{1-p}\right).
  \end{IEEEeqnarray*}
  Note that $p \log n \to 0$ and $t_n \to \infty$, so $\E[X] \to \infty$.
  Thus $P[X \leq \frac{1}{2} \E[X]] \to 0$. 
\end{IEEEproof}}

\newtheorem*{T1}{Theorem~\ref{thm:converse}}
\begin{T1}
  \thmconverse
\end{T1}

\begin{IEEEproof}
  We have $\frac{p_{11}p_{00}}{p_{10}p_{01}} > 1$, so from Lemma~\ref{lemma:posterior},
  if $\Delta(G_a,G_b) \geq \Delta(G_a, G_b \circ \tau)$, then the posterior probability of $\pi$ is at least as large as the true permutation. 
  From Lemma~\ref{lemma:intersection}, there are at least $|Aut(G_a \wedge G_b)|$ such permutations.
  Thus any estimator for $\Pi$ succeeds with probability at most $|Aut(G_a \wedge G_b)|^{-1}$.
  The graph $G_a \wedge G_b$ is distributed as $ER(n,p_{11})$.
  With high probability, the size of the automorphism group of an $ER(n,p_{11})$ graph goes to infinity with $n$.
  More precisely, if $p_{11} \leq \frac{\log n - \omega(1)}{n}$, then from Lemma~\ref{lemma:isolated} there is some sequence $\epsilon_n \to 0$ such that 
  \begin{equation*}
    P\left[|Aut(G_a \wedge G_b)|^{-1} \geq \epsilon_n\right] \leq \epsilon_n.
  \end{equation*}
  Any estimator succeeds with probability at most $2\epsilon_n$.
\end{IEEEproof}


\appendices
\section{Generating function proofs}
\label{app:gf}
Let $\sigma$ be a permutation of $[\ell]$ with a single cycle and let $x$ and $y$ be matrices of formal variables indexed by $[2] \times [2]$.
For $\ell \in \N$, define the generating function
\begin{equation*}
  b_{\ell}(x,y) = \sum_{g \in [2]^{[\ell]}} \sum_{h \in [2]^{[\ell]}} x^{\mu(g,h)} y^{\mu(g \circ \sigma ,h)} .
\end{equation*}

\begin{lemma}
  \label{lemma:AB-equiv}
  \begin{equation*}
    a_{\ell}\left(x \odot y, \frac{y_{01}y_{10}}{y_{00}y_{11}} \right) = b_{\ell}(x,y)
  \end{equation*}
\end{lemma}
\begin{IEEEproof}
  Define
  \begin{align*}
    a_{\ell,g,h}(w,z) &= w^{\mu(g,h)} z^{\delta(\sigma;g,h)}\\
    b_{\ell,g,h}(x,y) &= x^{\mu(g,h)} y^{\mu(g \circ \sigma, h)}.
  \end{align*}
  For each $g,h \in [2]^{[\ell]}$,
  \begin{align*}
    a_{\ell,g,h}\left(x \odot y, \frac{y_{01}y_{10}}{y_{00}y_{11}} \right)
    &= (x \odot y)^{\mu(g,h)} \left(\frac{y_{01}y_{10}}{y_{00}y_{11}} \right)^{\delta(\sigma;g,h)}\\
    &\eql{a} x^{\mu(g,h)} y^{\mu(g,h)} y^{\mu(g \circ \sigma, h) - \mu(g, h)}\\
    &= x^{\mu(g,h)} y^{\mu(g \circ \sigma, h)}\\
    &= b_{\ell,g,h}(x,y)
  \end{align*}
  where $(a)$ follows from from Lemma~\ref{lemma:type-diff}.
  We have
  \begin{align*}
    a_{\ell}(w,z) &= \sum_{g,h \in [2]^{[\ell]}} a_{\ell,g,h}(w,z)\\
    b_{\ell}(z,y) &= \sum_{g,h \in [2]^{[\ell]}} b_{\ell,g,h}(x,y)
  \end{align*}
  so the claimed identity follows.
\end{IEEEproof}

Let $x$ be a matrix of formal variables indexed by $[2] \times [2]$.
For $\ell \in \N$, define the generating function 
\begin{align}
  c_{\ell}(x) 
  &= \sum_{f \in [2]^\mathcal{S}} x^{\mu(f,f \circ \sigma)}. \label{def:c}
\end{align}

\begin{lemma}
  \label{lemma:gf-identity-one}
  For all $\ell \in \N$,
  \[
    b_{\ell}(x,y) = c_{\ell}(xy^{\top})
  \]
  where $y^{\top}$ is the transpose of $y$.
\end{lemma}
\begin{IEEEproof}
  Consider a cyclic $([2] \times [2])$-ary sequence $(g,h)$ indexed by $\mathcal{S}$, where $|\mathcal{S}| = l$, with the cyclic permutation $\sigma$.
  We have
  \begin{align*}
        \eq b_{\ell}(x,y) 
    &      = \sum_{g \in [2]^\mathcal{S}} \sum_{h \in [2]^\mathcal{S}} \prod_{e \in \mathcal{S}} x_{g(e),h(e)} \, y_{g(\sigma(e)),h(e)} \\
    &\eql{a} \sum_{g \in [2]^\mathcal{S}} \prod_{e \in \mathcal{S}} \sum_{h(e) \in [2]} x_{g(e),h(e)} \, y_{g(\sigma(e)),h(e)} \\
    &      = \sum_{g \in [2]^\mathcal{S}} \prod_{e \in \mathcal{S}} (xy^{\top})_{g(e),g(\sigma(e))} \\
    &      = c_{\ell}(xy^{\top})
  \end{align*}
  where $(a)$ follows because $h(e)$ appears in only one term of the product over $\mathcal{S}$.
  (In contrast $g(e)$ appears in both the $e$ term and the  $\sigma^{-1}(e)$ term.)
\end{IEEEproof}

\subsection{Cyclic sequence bijections}
Let $f$ be a cyclic $[2]$-ary sequences indexed by $\mathcal{S}$ with no consecutive ones (i.e. there is no $e \in \mathcal{S}$ such that $f(e) = 1$ and $f(\sigma(e)) = 1$).
Each such $f$ corresponds to a partition of $\mathcal{S}$ into blocks of size one and two: $e \in \mathcal{S}$ is in the same block as $\sigma(e)$ when $f(e) = 1$.
Thus each block either contains a one followed by a zero or just a zero.
Call this a cyclic partition of $\mathcal{S}$.

There are $\mu(f,f \circ \sigma)_{00}$ blocks of size one and $\mu(f,f \circ \sigma)_{01} = \mu(f,f \circ \sigma)_{10}$ blocks of size two.
Define the following generating function for these restricted cyclic sequences:
\[
  d_{\ell}(u,v) = \sum_{f \in [2]^\mathcal{S} : \mu(f,f \circ \sigma)_{11} = 0} u^{\mu(f,f \circ \sigma)_{00}} v^{\mu(f,f \circ \sigma)_{01}}.
\]

\begin{lemma}
  \label{lemma:bijection-one}
  \[
    c_{\ell}(x) = d_{\ell}(x_{00} + x_{11}, x_{01}x_{10} - x_{00}x_{11})
  \]
\end{lemma}
\begin{IEEEproof}
  The left side of the equation enumerates cyclic $[2]$-ary sequences as described in \eqref{def:c}.
  From each such sequence, we can obtain a cyclic partition of $\mathcal{S}$ as follows.
  If $f(e) = 0$ and $f(\sigma(e)) = 1$, put $e$ and $\sigma(e)$ in a block of size two and tag the block with the formal variables $x_{01}x_{10}$.
  If $f(e) = 0$ and $f(\sigma(e)) = 0$, put $e$ in a block of size one and tag the block with $x_{00}$.
  If $f(e) = 1$ and $f(\sigma(e)) = 1$, put $\sigma(e)$ in a block of size one and tag the block with $x_{11}$.

  The right side enumerates cyclically-partitioned cyclic $[2]$-ary sequences built of the following blocks: $(0,x_{00})$, $(1,x_{11})$, $(01,x_{01}x_{10})$, and $(01,-x_{00}x_{11})$.
  Let $f$ be a cyclic $[2]$-ary sequence with $k = \mu(f, f \circ \sigma)$.
  Then $f$ can be partitioned in $3^{k_{01}}$ ways: each appearance of $01$ can be produced by $(0,x_{00})$ followed by $(1,x_{11})$, by $(01,x_{01}x_{10})$, and by $(01,-x_{00}x_{11})$.
  Thus the total contribution of the partitions of $f$ to the right hand side is
\[
  (x_{00}x_{11} + x_{01}x_{10} - x_{00}x_{11})^{k_{10}}x_{00}^{k_{00}}x_{11}^{k_{11}}
  = x_{01}^{k_{01}}x_{10}^{k_{10}}x_{00}^{k_{00}}x_{11}^{k_{11}}.
\]
That is, only the partition that is counted on the left side is not canceled by some other partition.
  Figure~\ref{figure:cancellation} illustrates the partitions compatible with one example of $f$.
\end{IEEEproof}

\begin{figure}
\centering
 \begin{tikzpicture}[
 scale = 1/3,
 text height = 1.5ex,
 text depth =.1ex,
 b/.style={very thick}]

 \foreach \j in {0,1.5,3,4.5,6,7.5,9,10.5,12}
 \foreach \i in {0,1,3}
 \draw (\i + 0.5, \j + 0.5) node {$1$};

 \foreach \j in {0,1.5,3,4.5,6,7.5,9,10.5,12}
 \foreach \i in {2,4,5}
 \draw (\i + 0.5, \j + 0.5) node {$0$};

 \foreach \j in {0,1.5,3,4.5,6,7.5,9,10.5,12}
 \draw[b] (0,\j) -- (6,\j);

 \foreach \j in {0,1.5,3,4.5,6,7.5,9,10.5,12}
 \draw[b] (0,\j+1) -- (6,\j+1);

 \foreach \i in {0,1,2,3,4,5,6}
 \draw[b] (\i,0) -- (\i,1);
 
 \foreach \i in {1,2,3,4,5}
 \draw[b] (\i,1.5) -- (\i,2.5);
 
 \foreach \i in {1,2,3,4,5}
 \draw[b] (\i,3) -- (\i,4);
 
 \foreach \i in {0,1,2,4,5,6}
 \draw[b] (\i,4.5) -- (\i,5.5);
 
 \foreach \i in {0,1,2,4,5,6}
 \draw[b] (\i,6) -- (\i,7);
 
 \foreach \i in {1,2,4,5}
 \draw[b] (\i,7.5) -- (\i,8.5);
 
 \foreach \i in {1,2,4,5}
 \draw[b] (\i,9) -- (\i,10);
 
 \foreach \i in {1,2,4,5}
 \draw[b] (\i,10.5) -- (\i,11.5);
 
 \foreach \i in {1,2,4,5}
 \draw[b] (\i,12) -- (\i,13);

 \draw (3,5) node {$\bullet$};
 \draw (6,2) node {$\bullet$};
 \draw (3,8) node {$\bullet$};
 \draw (6,8) node {$\bullet$};
 \draw (3,11) node {$\bullet$};
 \draw (6,9.5) node {$\bullet$};

 \draw (16,12.5) node {$x_{00}x_{01}^2x_{10}^2x_{11}$};
 \draw (16,11)   node {$-x_{00}^2x_{01}x_{10}x_{11}^2$};
 \draw (16,9.5)  node {$-x_{00}^2x_{01}x_{10}x_{11}^2$};
 \draw (16,8)    node {$x_{00}^3x_{11}^3$};
 \draw (16,6.5)  node {$x_{00}^2x_{01}x_{10}x_{11}^2$};
 \draw (16,5)    node {$-x_{00}^3x_{11}^3$};
 \draw (16,3.5)  node {$x_{00}^2x_{01}x_{10}x_{11}^2$};
 \draw (16,2)    node {$-x_{00}^3x_{11}^3$};
 \draw (16,0.5)  node {$x_{00}^3x_{11}^3$};

 \end{tikzpicture}
 \caption{An illustration of the cancellations that occur on the right hand side of the identity in Lemma~\ref{lemma:bijection-one}. There are nine cyclic partitions compatible with the labeling $110100$. Blocks containing $01$ are tagged with $-x_{00}x_{11}$ if they are marked with $\bullet$ and are tagged with $x_{01}x_{10}$ otherwise.
 The first row contains the only partition that produces the correct monomial for the labeling.}
\label{figure:cancellation}
\end{figure}


The previous generating function identities of this section combine to give the following theorem.
\begin{theorem}
  \label{thm:main-identity}
  \bcomment{
  Let $\tau$ be a permutation of $\mathcal{S}$ with $t_{\ell}$ cycles of length $\ell$ for each $\ell \in \N$.
  Then
  \[
    A_{\mathcal{S},\tau}(w,z) = \prod_{\ell \in \N} \left( 2 \sum_i \binom{\ell}{2i} \left(\frac{u}{2}\right)^{\ell-2i} \left(\frac{u^2}{4} + v\right)^i\right)^{t_{\ell}} 
  \]}
For all $\ell \in \N$,
\[
  a_{\ell}(w,z) = d_{\ell}(u,v)
\]
where
  \begin{align*}
    u &= w_{00}+w_{01}+w_{10}+w_{11} \\
    v &= w_{00}w_{11} (z - 1) + w_{01}w_{10} (z^{-1} - 1)
  \end{align*}
\end{theorem}
\begin{IEEEproof}
  For $i,j \in [2]$, let $w_{ij} = x_{ij}y_{ij}$.
  Let $z = \frac{y_{01}y_{10}}{y_{00}y_{11}}$.
  Then from Lemma~\ref{lemma:AB-equiv} we have $a_{\ell}(w,z) = b_{\ell}(x,y)$.  
  Combining Lemmas~\ref{lemma:cycle-decomp}, \ref{lemma:gf-identity-one}, and \ref{lemma:bijection-one}, we have
  \[
    b_{\ell} (x,y) = d_{\ell}(\operatorname{tr}(xy^{\top}), -\det(xy^{\top}))^{t_{\ell}}.
  \]
  Then 
  \begin{align*}
    &\eq \det(xy^{\top})\\
    &= (x_{00}x_{11} - x_{01}x_{10})(y_{00}y_{11} - y_{01}y_{10})\\
    &= w_{00}w_{11} + w_{01}w_{10} - x_{01}x_{10}y_{00}y_{11} - x_{00}x_{11}y_{01}y_{10}\\
    &= w_{00}w_{11} + w_{01}w_{10} - w_{01}w_{10}z^{-1} - w_{00}w_{11}z
  \end{align*}
  and $\operatorname{tr}(xy^{\top}) = u$.
\end{IEEEproof}

Finally, we present a simple expression for $d_{\ell}$.
\begin{lemma}
  \label{lemma:bijection-two}
  \[
    d_{\ell}(2u,v) = 2 \sum_i \binom{\ell}{2i} u^{\ell-2i} (u^2 + v)^i 
  \]
\end{lemma}
\begin{IEEEproof}
  Both sides enumerates cyclically-partitioned cyclic $[2]$-ary sequences built of the following blocks: $(0,u)$, $(1,u)$, $(01,v)$.
  On the left side, the $[2]$-ary labels only serve to distinguish the two types of blocks of size one.
  The right side enumerates each cyclic $[2]$-ary sequence $f$ via its cyclic sequence of differences, $g: e \mapsto f(\sigma(e))-f(e)$.
  Each $g$ has the same number of ones as negative ones and there are $2 \binom{\ell}{2i}$ sequences with $i$ ones and $i$ negative ones.
  The ones in $g$ correspond to appearances of $01$ in $f$, which can be produced either by the block $(0,u)$ followed by the block $(1,u)$ or by the block $(01,v)$.
  The other $l-2i$ positions in $f$ are produced by either $(0,u)$ or $(1,u)$.
\end{IEEEproof}

\subsection{Generating function inequalities}
\begin{lemma}
  \label{lemma:l-cycle-two-cycle}
  Let $u,v \in \R$ such that $u \geq 0$ and $u^2+4v \geq 0$.  
  Then for all $\ell \geq 2$, $d_{\ell}(u,v) \leq d_2(u,v)^{\ell/2}$.
\end{lemma}
\begin{IEEEproof}
    We have
  \begin{align*}
      d_{\ell}(u,v) 
    &\eql{a} \sum_i \binom{\ell}{2i} 2 \left(\frac{u}{2}\right)^{\ell-2i} \left(\frac{u^2}{4}+v\right)^i \\
    &\eql{b} \sum_j \binom{\ell}{j} (1 + (-1)^j) \left(\frac{u}{2}\right)^{\ell-j} \left(\frac{u^2}{4}+v\right)^{\frac{j}{2}} \\
    &=       \sum_j \binom{\ell}{j} \left(\frac{u}{2}\right)^{\ell-j} \left(\sqrt{\frac{u^2}{4}+v}\right)^j \\
    &\eq +\sum_j \binom{\ell}{j} \left(\frac{u}{2}\right)^{\ell-j} \left(-\sqrt{\frac{u^2}{4}+v}\right)^j \\
    &\eql{c} \left(\frac{u}{2} + \sqrt{\frac{u^2}{4}+v}\right)^{\ell} + \left(\frac{u}{2} - \sqrt{\frac{u^2}{4}+v}\right)^{\ell} \\
    &\leql{d} \left(\left(\frac{u}{2} + \sqrt{\frac{u^2}{4} + v}\right)^2 + \left(\frac{u}{2} - \sqrt{\frac{u^2}{4} + v}\right)^2\right)^{\ell/2}\\
    &= (u^2 + 2v)^{\ell/2} \\
    &= d_2(u,v)^{\ell/2} 
  \end{align*}
  Here $(a)$ uses Lemma~\ref{lemma:bijection-two} and $(b)$ uses by the binomial theorem.
  In $(b)$, we note that $i$ only appears as in the expression as $2i$, so we switch to a sum over $j$ with a factor of $(1 + (-1)^j)$ to eliminate the odd terms.
  In $(c)$, we apply the binomial theorem to each term.
  In $(d)$, we have used a standard $p$-norm inequality: for a vector $\bfx$, $\|\bfx\|_\ell \leq \|\bfx\|_2$ when $\ell \geq 2$.
\end{IEEEproof}

\newtheorem*{T4}{Theorem~\ref{thm:l-cycle-two-cycle}}
\begin{T4}
  \thmgf
\end{T4}
\begin{IEEEproof}
  From Theorem~\ref{thm:main-identity} we have $a_{\ell}(w,z) = d_{\ell}(u,v)$ where
  \begin{align*}
    u &= w_{00}+w_{01}+w_{10}+w_{11} \\
    v &= w_{00}w_{11} z - w_{00}w_{11} - w_{01}w_{10} + w_{01}w_{10} z^{-1}
  \end{align*}
  Note that for all $w \in \R_{> 0}^{[2] \times [2]}$, we have $\frac{w_{00}+w_{11}}{2} \geq \sqrt{w_{00}w_{11}}$ and $\frac{w_{01}+w_{10}}{2} \geq \sqrt{w_{01}w_{10}}$.
  Thus
  \begin{align*}
    \frac{u^2}{4} &\geq \frac{(w_{00}+w_{01}+w_{10}+w_{11})^2}{4}\\
    &\geq (\sqrt{w_{00}w_{11}} + \sqrt{w_{01}w_{10}})^2\\
    &\geq (\sqrt{w_{00}w_{11}} - \sqrt{w_{01}w_{10}})^2\\
    &\geql{a} -v
  \end{align*}
  where $(a)$ follows from \eqref{min-z} in Section~\ref{subsection:tail}.
  Thus the conditions of Lemma~\ref{lemma:l-cycle-two-cycle} are satisfied and the claimed inequality follows.
\end{IEEEproof}



\bibliographystyle{IEEEtran}
\bibliography{IEEEabrv,deanon}

\begin{thebibliography}{10}
\providecommand{\url}[1]{#1}
\csname url@samestyle\endcsname
\providecommand{\newblock}{\relax}
\providecommand{\bibinfo}[2]{#2}
\providecommand{\BIBentrySTDinterwordspacing}{\spaceskip=0pt\relax}
\providecommand{\BIBentryALTinterwordstretchfactor}{4}
\providecommand{\BIBentryALTinterwordspacing}{\spaceskip=\fontdimen2\font plus
\BIBentryALTinterwordstretchfactor\fontdimen3\font minus
  \fontdimen4\font\relax}
\providecommand{\BIBforeignlanguage}[2]{{%
\expandafter\ifx\csname l@#1\endcsname\relax
\typeout{** WARNING: IEEEtran.bst: No hyphenation pattern has been}%
\typeout{** loaded for the language `#1'. Using the pattern for}%
\typeout{** the default language instead.}%
\else
\language=\csname l@#1\endcsname
\fi
#2}}
\providecommand{\BIBdecl}{\relax}
\BIBdecl

\bibitem{cullina_improved_2016}
D.~Cullina and N.~Kiyavash, ``Improved achievability and converse bounds for
  {Erdos}-{Rényi} graph matching,'' in \emph{Proceedings of the 2016 {ACM}
  {SIGMETRICS} {International} {Conference} on {Measurement} and {Modeling} of
  {Computer} {Science}}.\hskip 1em plus 0.5em minus 0.4em\relax ACM, 2016, pp.
  63--72.

\bibitem{narayanan_de-anonymizing_2009}
A.~Narayanan and V.~Shmatikov, ``De-anonymizing social networks,'' in
  \emph{Security and {Privacy}, 2009 30th {IEEE} {Symposium} on}.\hskip 1em
  plus 0.5em minus 0.4em\relax IEEE, 2009, pp. 173--187.

\bibitem{pedarsani_privacy_2011}
P.~Pedarsani and M.~Grossglauser, ``On the privacy of anonymized networks,'' in
  \emph{Proceedings of the 17th {ACM} {SIGKDD} international conference on
  {Knowledge} discovery and data mining}.\hskip 1em plus 0.5em minus
  0.4em\relax ACM, 2011, pp. 1235--1243.

\bibitem{wright_graphs_1971}
E.~M. Wright, ``Graphs on unlabelled nodes with a given number of edges,''
  \emph{Acta Mathematica}, vol. 126, no.~1, pp. 1--9, 1971.

\bibitem{bollobas_distinguishing_1982}
B.~Bollob\'{a}s, ``Distinguishing {Vertices} of {Random} {Graphs},''
  \emph{North-Holland Mathematics Studies}, vol.~62, pp. 33--49, Jan. 1982.

\bibitem{onaran_optimal_2016}
E.~Onaran, S.~Garg, and E.~Erkip, ``Optimal de-anonymization in random graphs
  with community structure,'' in \emph{Signals, {Systems} and {Computers}, 2016
  50th {Asilomar} {Conference} on}.\hskip 1em plus 0.5em minus 0.4em\relax
  IEEE, 2016, pp. 709--713.

\bibitem{kazemi_when_2015}
E.~Kazemi, L.~Yartseva, and M.~Grossglauser, ``When {Can} {Two} {Unlabeled}
  {Networks} {Be} {Aligned} {Under} {Partial} {Overlap}?'' in \emph{Proceedings
  of the 53rd {Annual} {Allerton} {Conference} on {Communication}, {Control},
  and {Computing}}, 2015.

\bibitem{yartseva_performance_2013}
L.~Yartseva and M.~Grossglauser, ``On the performance of percolation graph
  matching,'' in \emph{Proceedings of the first {ACM} conference on {Online}
  social networks}.\hskip 1em plus 0.5em minus 0.4em\relax ACM, 2013, pp.
  119--130.

\bibitem{kazemi_growing_2015}
E.~Kazemi, S.~H. Hassani, and M.~Grossglauser, ``Growing a {Graph} {Matching}
  from a {Handful} of {Seeds},'' in \emph{Proceedings of the {Vldb} {Endowment}
  {International} {Conference} on {Very} {Large} {Data} {Bases}}, vol.~8, 2015.

\bibitem{shirani_seeded_2017}
F.~Shirani, S.~Garg, and E.~Erkip, ``Seeded {Graph} {Matching}: {Efficient}
  {Algorithms} and {Theoretical} {Guarantees},'' \emph{arXiv preprint
  arXiv:1711.10360}, 2017.

\bibitem{ji_structural_2014}
S.~Ji, W.~Li, M.~Srivatsa, and R.~Beyah, ``Structural {Data}
  {De}-anonymization: {Quantification}, {Practice}, and {Implications},'' in
  \emph{Proceedings of the 2014 {ACM} {SIGSAC} {Conference} on {Computer} and
  {Communications} {Security}}.\hskip 1em plus 0.5em minus 0.4em\relax ACM,
  2014, pp. 1040--1053.

\bibitem{ji_your_2015}
S.~Ji, W.~Li, N.~Z. Gong, P.~Mittal, and R.~Beyah, ``On {Your} {Social}
  {Network} {De}-anonymizablity: {Quantification} and {Large} {Scale}
  {Evaluation} with {Seed} {Knowledge},'' 2015.

\bibitem{chvatal_tail_1979}
V.~Chv\'{a}tal, ``The tail of the hypergeometric distribution,'' \emph{Discrete
  Mathematics}, vol.~25, no.~3, pp. 285--287, 1979.

\bibitem{bollobas_random_1998}
B.~Bollob\'{a}s, \emph{Random graphs}.\hskip 1em plus 0.5em minus 0.4em\relax
  Springer, 1998.

\end{thebibliography}

\end{document}